**SARS-CoV-2 Wastewater Genomic Surveillance: Approaches, Challenges, and Opportunities**

**Viorel Munteanu***
Department of Computers, Informatics and Microelectronics, Technical University of Moldova, Chisinau, 2045, Moldova
Department of Biological and Morphofunctional Sciences, College of Medicine and Biological Sciences, Stefan cel Mare University of Suceava, 720229 Suceava, Romania
Email: viorel.munteanu@lt.utm.md
ORCID: https://orcid.org/0000-0002-4133-5945

**Michael A. Saldana***
Astani Department of Civil and Environmental Engineering, University of Southern California, 3620 South Vermont Avenue, Los Angeles, CA 90089, USA
Email: michael.saldana.0@usc.edu
ORCID: https://orcid.org/0009-0007-7253-9434

**David Dreifuss**
Department of Biosystems Science and Engineering, ETH Zurich, Basel, Switzerland, SIB Swiss Institute of Bioinformatics, Lausanne, Switzerland
Email: david.dreifuss@bsse.ethz.ch
ORCID: https://orcid.org/0000-0002-5827-5387

**Wenhao O. Ouyang**
Department of Biochemistry, University of Illinois at Urbana-Champaign, Urbana, IL 61801, USA
Email: wenhaoo2@illinois.edu
ORCID: https://orcid.org/0000-0002-3618-9808

**Jannatul Ferdous**
Department of Bioinformatics and Genomics, University of North Carolina, Charlotte, USA
Email: jferdous@charlotte.edu
ORCID: https://orcid.org/0000-0003-3053-9616

**Fatemeh Mohebbi**
Department of Clinical Pharmacy, School of Pharmacy, University of Southern California, 1540 Alcazar Street, Los Angeles, CA 90033, USA
Email: fmohebbi@usc.edu
ORCID: https://orcid.org/0000-0003-1395-1051

**Jessica Schlueter**
Department of Bioinformatics and Genomics, University of North Carolina, Charlotte, USA
Email: jschluet@charlotte.edu
ORCID:https://orcid.org/0000-0002-6490-0580

**Dumitru Ciorba**

Department of Computers, Informatics and Microelectronics, Technical University of Moldova, Chisinau, 2045, Moldova
Email: dumitru.ciorba@fcim.utm.md
ORCID: https://orcid.org/0000-0002-3157-5072

**Viorel Bostan**
Department of Computers, Informatics and Microelectronics, Technical University of Moldova, Chisinau, 2045, Moldova
Email: viorel.bostan@adm.utm.md
ORCID: https://orcid.org/0000-0002-2422-3538

**Victor Gordeev**
Department of Computers, Informatics, and Microelectronics, Technical University of Moldova, Chisinau, 2045, Republic of Moldova
Email: victor.gordeev@lt.utm.md
ORCID: https://orcid.org/0009-0005-1052-2552

**Justin Maine Su**
Titus Family Department of Clinical Pharmacy, USC Alfred E. Mann School of Pharmacy and Pharmaceutical Sciences, University of Southern California, Los Angeles, CA 90089, USA
Email: jmsu@usc.edu
ORCID: https://orcid.org/0000-0002-0912-4785

**Nadiia Kasianchuk**
Faculty of Biology, Adam Mickiewicz University Poznan, 61-712, Poland
Faculty of Pharmacy, Bogomolets National University, Kyiv, 01601, Ukraine
Kyiv School of Economics, Kyiv, 02000, Ukraine
Email: nadkas2@st.amu.edu.pl
ORCID: https://orcid.org/0000-0002-9732-329X

**Nitesh Kumar Sharma**
Titus Family Department of Clinical Pharmacy, USC Alfred E. Mann School of Pharmacy and Pharmaceutical Sciences, University of Southern California, 1540 Alcazar Street, Los Angeles, CA 90033, USA
Email: prince26121991@gmail.com
ORCID: https://orcid.org/0000-0001-6379-1426

**Sergey Knyazev**
Department of Pathology and Laboratory Medicine, David Geffen School of Medicine, University of California, Los Angeles, Los Angeles, CA, USA
Titus Family Department of Clinical Pharmacy, USC Alfred E. Mann School of Pharmacy and Pharmaceutical Sciences, University of Southern California, 1540 Alcazar Street, Los Angeles, CA 90033, USA
Email: sergey.n.knyazev@gmail.com
ORCID:https://orcid.org/0000-0003-0385-1831

**Eva Aßmann**
Genome Competence Center (MF 1), Method Development and Research Infrastructure, Robert Koch Institute, 13353 Berlin, Germany
Centre for Artificial Intelligence in Public Health Research (ZKI-PH), Robert Koch Institute, 13353 Berlin, Germany
Email: AssmannE@rki.de
ORCID: 0000-0002-7249-069X

**Andrei Lobiuc**
College of Medicine and Biological Sciences, University of Suceava, Suceava, 720229, Romania
Email: andrei.lobiuc@usm.ro
ORCID: https://orcid.org/0000-0001-5854-8261

**Mihai Covasa**
College of Medicine and Biological Sciences, University of Suceava, Suceava, 720229, Romania
College of Osteopathic Medicine of the Pacific, Western University of Health Sciences, Pomona, CA 91766, USA
Email: mcovasa@usm.ro
ORCID: https://orcid.org/0000-0002-6266-4457

**Keith A. Crandall**
Computational Biology Institute, Department of Biostatistics and Bioinformatics, Milken Institute
School of Public Health, George Washington University, Washington, DC 20052, USA
Email: kcrandall@gwu.edu
ORCID: https://orcid.org/0000-0002-0836-3389

**Nicholas C. Wu**
Department of Biochemistry, University of Illinois at Urbana-Champaign, Urbana, IL 61801, USA
Centre for Biophysics and Computational Biology, University of Illinois at Urbana-Champaign, Urbana, IL 61801, USA
Carl R. Woese Institute for Genomic Biology, University of Illinois at Urbana-Champaign, Urbana, IL 61801, USA
Carle Illinois College of Medicine, University of Illinois at Urbana-Champaign, Urbana, IL 61801, USA
Email: nicwu@illinois.edu
ORCID: https://orcid.org/0000-0002-9078-6697

**Christopher E. Mason**
Department of Physiology and Biophysics, Weill Cornell Medicine, New York, NY, USA, 10065
Email: chm2042@med.cornell.edu
ORCID: https://orcid.org/0000-0002-1850-1642

**Braden T Tierney**

Department of Physiology and Biophysics, Weill Cornell Medicine, New York, NY, USA, 10065
Email: btt4001@med.cornell.edu
ORCID: https://orcid.org/0000-0002-7533-8802

**Alexander G Lucaci**
Department of Physiology and Biophysics, Weill Cornell Medicine, New York, NY, USA, 10065
Email: agl4001@med.cornell.edu
ORCID: https://orcid.org/0000-0002-4896-6088

**Roel A. Ophoff**
Center for Neurobehavioral Genetics, Semel Institute for Neuroscience and Human Behavior, University of California Los Angeles, Los Angeles, CA, USA.
Email: ophoff@ucla.edu
ORCID: https://orcid.org/0000-0002-8287-6457

**Cynthia Gibas**
Department of Bioinformatics and Genomics, University of North Carolina, Charlotte, USA
Email: cgibas@charlotte.edu
ORCID: https://orcid.org/0000-0002-1288-9543

**Piotr Rzymski**
Department of Environmental Medicine,
Poznan University of Medical Sciences,
Poznan, Poland
E-mail: rzymskipiotr@ump.edu.pl
ORCID: https://orcid.org/0000-0002-4713-0801

**Pavel Skums**
School of Computing, University of Connecticut, Storrs, Connecticut
Email: pavel.skums@uconn.edu
ORCID: https://orcid.org/0000-0003-4007-5624

**Helena Solo-Gabriele**
Department of Chemical, Environmental, and Materials Engineering
University of Miami, Coral Gables, FL USA
Email: hmsolo@miami.edu
ORCID: https://orcid.org/0000-0003-3390-3823

**Beerenwinkel Niko**
Department of Biosystems Science and Engineering, ETH Zurich, Basel, Switzerland, SIB Swiss Institute of Bioinformatics, Lausanne, Switzerland
Email: niko.beerenwinkel@bsse.ethz.ch
ORCID: https://orcid.org/0000-0002-0573-6119

**Alex Zelikovsky**
Department of Computer Science, College of Art and Science, Georgia State University, Atlanta, GA, USA
Email: alex.zelikovsky@gmail.com
ORCID: https://orcid.org/0000-0003-4424-4691

**Martin Hölzer**[†]
Genome Competence Center (MF1), Method Development and Research Infrastructure, Robert Koch Institute, 13353 Berlin, Germany
Email:HoelzerM@rki.de
ORCID: 0000-0001-7090-8717

**Adam Smith**[†]
Astani Department of Civil and Environmental Engineering, University of Southern California, 920 Downey Way; BHE 221 Los Angeles, CA 90089
Email: smithada@usc.edu
ORCID: https://orcid.org/0000-0002-3964-7544

**Serghei Mangul**[†‡]
Department of Clinical Pharmacy, Alfred E. Mann School of Pharmacy and Pharmaceutical Sciences, University of Southern California, Los Angeles, CA 90033, USA
Department of Biological and Morphofunctional Sciences, College of Medicine and Biological Sciences, Stefan cel Mare University of Suceava, 720229 Suceava, Romania
serghei.mangul@gmail.com
ORCID: 0000-0003-4770-3443

* These authors contributed equally to this work
[†] These authors jointly supervised this work
[‡] Corresponding author: serghei.mangul@gmail.com

## Abstract

During the SARS-CoV-2 pandemic, wastewater-based genomic surveillance (WWGS) emerged as an efficient virus surveillance tool capable of accounting for asymptomatic cases, identifying known and novel mutations, and assigning them to lineages and sublineages of the virus. WWGS also offers the potential to uncover novel or cryptic lineages, although identifying and defining these lineages solely from wastewater presents challenges. Besides the significant advantages of WWGS for monitoring SARS-CoV-2 viral spread, technical challenges remain, including coverage dropouts and poor quality of wastewater sequencing (WWS) data due to viral RNA fragmentation and degradation. Subsequently, WWGS analysis requires advanced computational tools that are yet to be developed and benchmarked. The existing bioinformatics tools used to analyze WWS data are often based on previously developed computational methods for variant calling, *de-novo* assembly, or quantifying the expression of transcripts. Those methods were not specifically developed for WWS data and must be optimized to address unique challenges associated with wastewater analysis. While specialized tools for analyzing WWS data have also been developed recently, it remains to be seen how they will perform given the ongoing evolution of SARS-CoV-2 and the decline in testing and patient-based genomic surveillance. Here, we explore opportunities and challenges associated with WWGS, including aspects such as sample preparation, sequencing technology, and bioinformatics methods. We also highlight its potential role in monitoring infectious diseases and informing public health responses by complementing other surveillance systems. We also critically examine the practical applications and ethical considerations of WWGS, emphasizing the need for a robust framework and its translation into sustainable wastewater surveillance programs to fully realize its potential for public health.

## Introduction

Early in the COVID-19 pandemic, the presence of SARS-CoV-2 RNA in the feces of individuals infected with the virus, including those who are asymptomatic or have recovered from respiratory symptoms[1–5], prompted researchers to explore the use of wastewater networks for community-wide surveillance of SARS-CoV-2 prevalence. From April to July of 2020, several teams submitted proof-of-concept findings to peer-reviewed publications outlining the use and potential of wastewater-based genomic surveillance (WWGS) for SARS-CoV-2[6–16]. The remarkably rapid dissemination of methods and results during that period facilitated the widespread adoption of WWGS as a valuable tool for tracking the pandemic in municipal settings worldwide[6,10,11,17,18]. These accomplishments have emphasized the potential of wastewater testing for viral surveillance as a method to evaluate disease prevalence within the community and demonstrated that WWGS for SARS-CoV-2 can detect emerging lineages at an earlier stage compared to clinical monitoring[12,13,19]. In contrast to clinical samples, wastewater sampling allows the development of community-level profiles for SARS-CoV-2 loads encompassing positive and asymptomatic tested cases and asymptomatic and symptomatic non-tested cases. This approach has also demonstrated its feasibility in monitoring variants of concern (VOC) as designated by the World Health Organization[6,10,11,16,17], and can serve as a valuable warning system for detecting regional spikes for VOC[20–22]. In turn, this has assisted in developing new laboratory and bioinformatics methods to obtain and analyze wastewater sequencing (WWS) data. Although many laboratory methods and bioinformatics tools have been rapidly developed in response to the COVID-19 pandemic, ongoing efforts persist in advancing WWGS approaches. These endeavors aim to harness the potential of wastewater analysis for monitoring and detecting viral genetic material, thereby offering valuable insights and enhancing our understanding of the pandemic's spread and dynamics. Wastewater-based monitoring of SARS-CoV-2 epidemiology has demonstrated its efficacy in tracking SARS-CoV-2 viral infection dynamics in numerous countries around the globe[6,10,11,16,23]. WWGS can serve many purposes, from early detection of isolated outbreaks and allocation of resources to hot spot areas in small and medium-sized populations (e.g., in an elderly home or on a campus) to large-scale surveillance and monitoring in large populations[24]. Wastewater became a promising core component of infectious disease monitoring, providing a variant-specific, community-representative picture of public health trends that can capture previously undetected spread and pathogen transmission links. Building on recent laboratory and analytical advances to identify the diverse pathogens present in wastewater will be essential for ongoing efforts to understand disease risks and will transform infectious disease surveillance[25].

Importantly, wastewater-based surveillance has been shown to provide balanced estimates of viral prevalence rates and does not require patient interaction, and can monitor entire communities, including underserved and vulnerable populations and asymptomatic cases[26–29]. SARS-CoV-2 WWGS can detect mutation patterns of virus variants earlier than clinical monitoring[12,13,19]. Additionally, it allows for the detection of novel cryptic variants, including those resistant to naturally acquired or vaccine-induced immunity, those rarely observed in clinical samples, and those from unsampled individuals with COVID-19 infections[23]. In contrast to clinical samples, wastewater sampling allows the development of community-level profiles encompassing positive, non-reporting, and asymptomatic viral loads. This non-invasive technique allows for analyzing a community within a given sewershed and can provide insight into rising mutations and potential variants of emerging concern[30].

Typical WWGS comprises multiple steps (Figure 1) after the initial assay design, including: wastewater sampling, viral particle concentration, and RNA extraction; SARS-CoV-2 targeting quantification; library preparation and sequencing; bioinformatics analysis, data sharing, and investigation of emerging mutations and lineages hinting towards potential outbreaks. WWGS involves a multitude of experimental and computational methods, offering researchers a wide array of options. However, this versatility is constrained by the inherent complexities of wastewater samples, including wastewater biophysicochemical properties, low viral loads, fragmented RNA, multiple genotypes, and a high background genomic noise. Despite current advancements, WWGS faces critical limitations, primarily due to potential experimental biases and the rigorous demands of computational analyses and interpretations.

Here, we present a comprehensive overview that delves into best practices, challenges, and opportunities surrounding WWGS for SARS-CoV-2 by providing a thorough examination of the current status of WWGS, shedding light on the obstacles and prospects with both experimental and bioinformatics methodologies. We thoroughly evaluate the available options and address the common challenges at each WWGS step. We first outline the critical components and approaches essential for effective WWGS. This includes an in-depth look at wastewater sampling techniques, virus concentration, RNA extraction methods, and the quantification methods critical to obtaining high-quality sequencing data and subsequent analysis. We also delve into the diverse sequencing technologies currently in use. Further, our review focuses on the bioinformatics analysis of WWS data. Here, we explore the indispensable tools for data quality control and error correction, the options for *de-novo* genome assembly, read mapping, single nucleotide variant (SNV) calling, as well as lineage calling and abundance estimation. Finally, we discuss the practical applications of WWGS. Our review covers its role in outbreak investigation, monitoring viral evolution and detecting novel mutations, tracing emerging lineages, and monitoring infection rates. Additionally, we examine how WWGS can assess the effectiveness of vaccinations, providing a crucial tool in public health decision-making. However, our review also acknowledges the challenges in WWGS, highlighting the technical and logistical hurdles that must be overcome to maximize its utility. Additionally, we address the ethical and privacy concerns associated with this surveillance method, emphasizing the need for careful consideration and management of these issues to responsibly harness the power of WWGS. The ultimate goal of this review is to motivate further advances in the field of WWGS, which has the significant potential to complement other surveillance systems and help guide public health strategies during various infectious diseases.

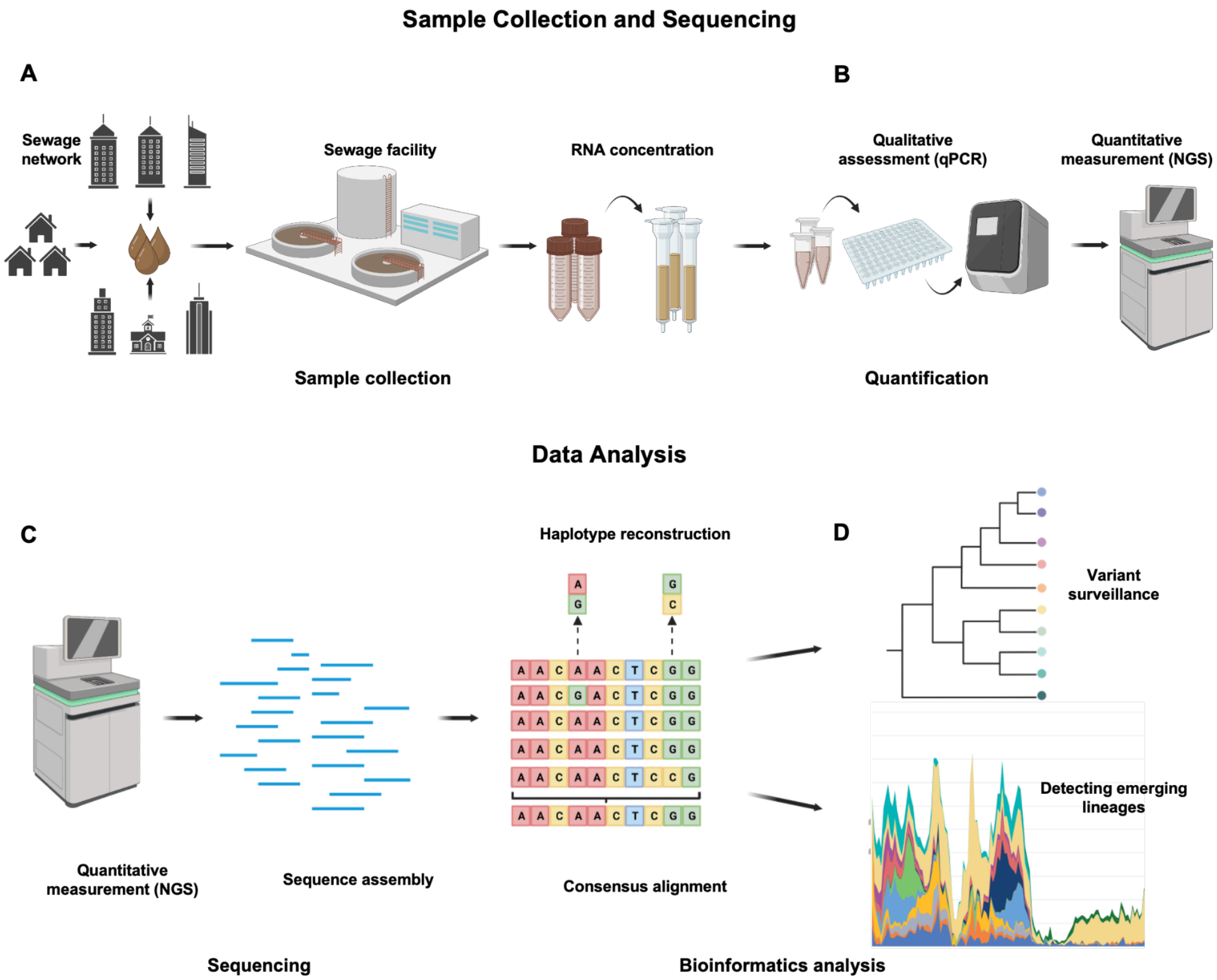

**Figure 1.** Overview of wastewater-based genomic surveillance (WWGS). (A) Workflow of sample collection and preparation for sequencing. Wastewater samples are collected from water reclamation facilities, followed by subsequent concentration and extraction of viral RNA. It is important to emphasize that wastewater for genomic surveillance can be collected and analyzed at different population levels, ranging from small (like a retirement home) to medium (like a small town) to large (like a county). The results need to be interpreted accordingly and put into context. (B) SARS-CoV-2 quantification using primers targeting different SARS-CoV-2 viral genes (such as N1, N2, and E-gene) to assess SARS-CoV-2 genome copy numbers quantitatively. Positive samples then proceed through library preparation and next-generation sequencing (NGS) technologies, usually via amplicon sequencing. (C) Data analysis pipeline of wastewater sequencing results. NGS reads are mapped to reference sequence and variant calling is performed. (D) Further, subsequent analysis contributes to variant surveillance, detection of emerging lineages, and identifying hot spot areas. The results need to be related to the size of the sampled population, integrated with data from other monitoring systems, and interpreted accordingly to be of full benefit to public health.

Although a commonly used qPCR-based approach for SARS-CoV-2 can reveal temporal changes in virus prevalence in a given population,[31–33] the nature of qPCR restricts its ability to detect existing and emerging SARS-CoV-2 variants and estimate their prevalence in the population. Next-generation sequencing (NGS) of wastewater samples can address these limitations, as an essential public health tool, offering an understanding of strain distribution, identifying emerging lineages and outbreaks in small populations, and tracking variants, especially in areas with limited clinical testing. The polymerase chain reaction (PCR) based methods, namely RT-qPCR, and more recent technologies such as RT-digital droplet PCR (RT-ddPCR), are relatively inexpensive and well-established and allow for the direct quantification of SARS-CoV-2 in wastewater samples, presenting the following advantages: an ability to probe a sample site at high frequency to generate real-time information; ease of implementation by any lab running standard PCR assays; short turn-around time, and lower costs of reagents[34]. As with any PCR assay development, methods and results must be carefully scrutinized to minimize the chance of false positives or over-interpretation. The genomic sequence targets of RT-qPCR/RT-ddPCR methods are also limited by fluorophores and the detection instrument[35]. Most critically, these PCR methods lag in discovering the emergence of new variants because they require a specific primer-probe design according to the details of the genomic information of new variants[36], usually derived from sequencing and analyzing patient samples. Thus, PCR is not effective for detecting new variants as they evolve. PCR-based techniques are limited to detecting and quantifying only known variants circulating in communities[35].

High-throughput sequencing can be employed to overcome the limitation of pre-defined sequence targets and to identify emerging viral lineages[35,37]. The use of sequencing technologies coupled with advanced bioinformatics methods for analyzing WWS data has provided unparalleled detail in assessing wastewater samples. Sequencing overcomes some of the limitations of PCR-based technologies, allowing for the comprehensive detection of SARS-CoV-2 mutation profiles present in wastewater samples. However, the tiling amplicon sequencing methods primarily used in SARS-CoV-2 surveillance are still somewhat vulnerable to unexpected changes in primer binding sequences as new lineages emerge. Sequence data collected at sufficient depth can be deconvoluted to estimate lineage and sublineage proportions. Including high-throughput sequencing with appropriate bioinformatics methods is the foundation of fundamental transformations of environmental genomic surveillance and virology to complement epidemiological data analysis and thus contribute to outbreak early detection and prevention[38–42].

To unlock the potential of WWS data, robust and accurate bioinformatics algorithms and analytical pipelines are needed. Additionally, comprehensive methodologies must be established to efficiently access SARS-CoV-2 viral genomic material, optimize adaptive sampling strategies, recover viral particles, and select appropriate sequencing technologies. Establishing such efforts is critical for the widespread adoption of WWGS as an all-encompassing approach for monitoring SARS-CoV-2 variant prevalence and detecting novel cryptic lineages. Overall, the true power of real-time SARS-CoV-2 tracking through WWGS comes from combining the two methodologies, qPCR and sequencing. By including sequencing approaches, samples can be explored for novel mutations and emerging lineages. When a concerning mutation profile or a new potential lineage are discovered, primers and probes can be adjusted for these new lineages to provide rapid turnaround monitoring via qPCR.

**Experimental approaches for effective wastewater genomic surveillance of viral RNA**

Access to SARS-CoV-2 viral genomic material in wastewater infrastructure is provided through a highly variable and complex wastewater collection system rather than direct access to individual clinical specimens. Ambient conditions within the wastewater collection system are harsh to viral material because of changing chemistry and physicochemical conditions outside the human host. Additionally, ambient conditions may include non-ideal and fluctuating temperatures, variable pH, water quality parameters (e.g., presence of DNases and RNases) that promote the degradation of the viral capsid and nucleic acids, and extended time from release from the human host to wastewater reclamation facilities (WRFs)[43,44]. As a result, viral genetic material can be severely degraded and fragmented prior to sample collection. Before collection, SARS-CoV-2 viruses may travel through the sewer network for several days; however, in untreated wastewater, the SARS-CoV-2 virus can survive for up to 10 days at room temperature (below 37°C) and between 30 and 60 days at 4°C[45]. Several studies have taken different approaches to overcoming the challenges presented by wastewater for WWGS (Table 1).

***Wastewater Sampling***

Outside of the host cell, viruses cannot replicate. As a result, the concentration of viral RNA, discharged into the wastewater collection system over time, can be monitored to reflect the population within a sewershed[49], but can also be biased by external factors such as heavy rainfall[47]. Thus, it is important to consider when and where sampling will occur, for this can dictate the level of RNA dilution and degradation of SARS-CoV-2[48]. Sampling techniques include grab samples at peak flow times (typically occurring between 0800-1100)[7,38,39] and 24-hour time-weighted composite samples using refrigerated autosamplers[7,40–42] (Figure 2).

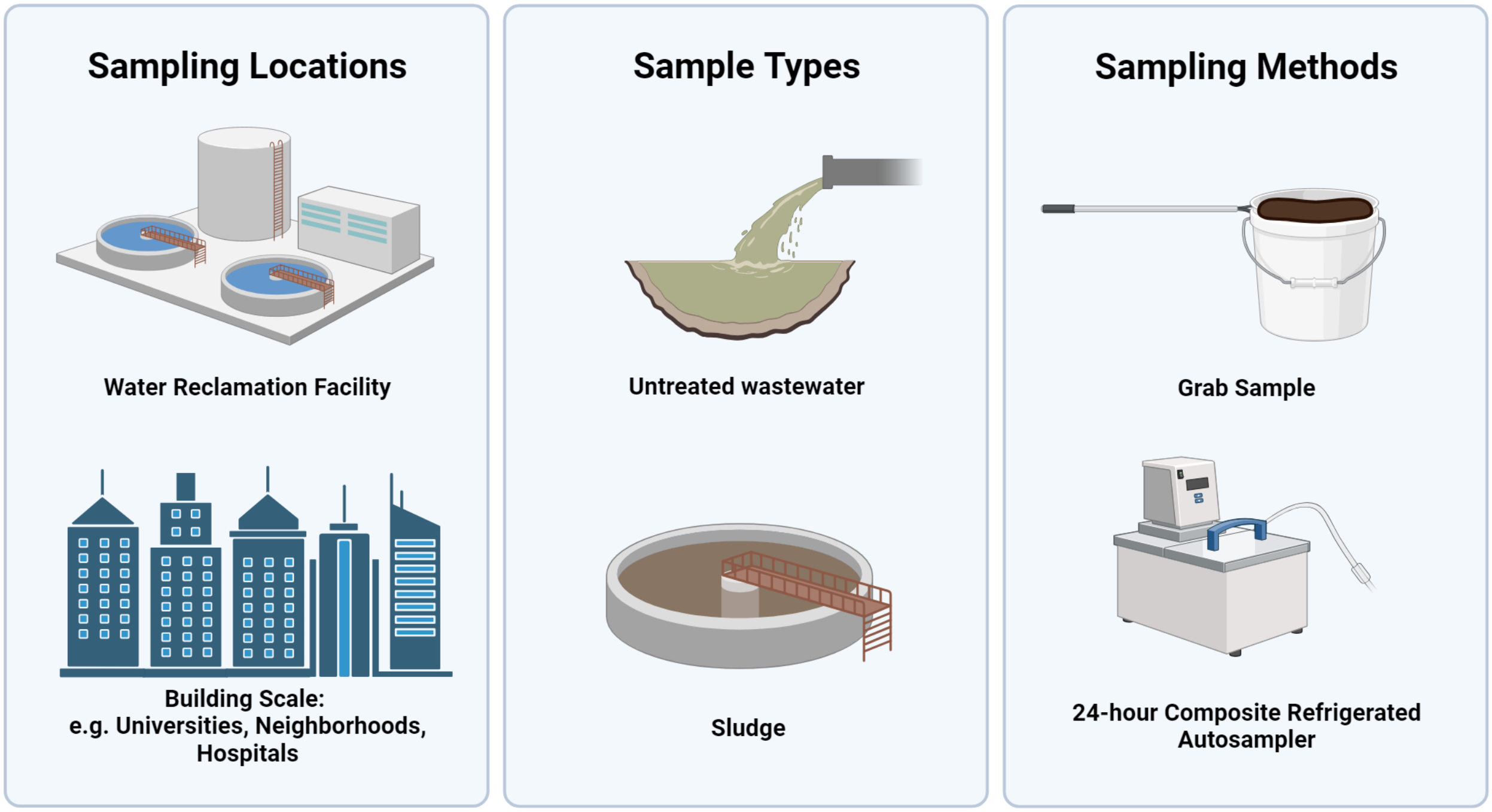

**Figure 2**: An outline of different sampling types and locations for WWGS. Genomic wastewater surveillance can be conducted across various population scales, from smaller settings such as retirement homes, to medium-sized communities, like small towns, and up to larger areas, like counties. The interpretation of the results must be done with respect to these varying scales and contextualized appropriately.

Wastewater sampling frequency varies across WWGS programs, e.g., ranging from once per week to daily. Clinical sampling data early during the pandemic was available daily and transitioned to a weekly basis later. One challenge in establishing relationships between wastewater SARS-CoV-2 RNA levels and disease prevalence is temporally matching the wastewater and clinical data. Data alignment necessitates aggregation so that both wastewater and clinical data are on the same time scale (e.g., weekly). In addition, the development of accurate models will also require an understanding of the progression of the disease and viral shedding for infected individuals. Moving averages (e.g., 7 days to 3 weeks) are also typically utilized to evaluate overall trends to reduce the short-term variability inherent in wastewater measurements and clinical case counts.[49]

The placement of sampling locations depends on the scale of SARS-CoV-2 monitoring. Collection at WRFs allows for monitoring SARS-CoV-2 from a potentially large population within the sewershed. Here, two different sample types can be collected: untreated wastewater and primary sludge (Figure 2). It has been demonstrated that primary sludge can provide higher sensitivity and less variance when compared to untreated wastewater[50]; however, primary sludge does not possess the same predictive capabilities as untreated wastewater, providing a much shorter lead-time to clinical diagnosis[12]. SARS-CoV-2 concentrations in untreated wastewater precede clinical data by 4-10 days[51]. It is important to note that the size of the surveyed population when collecting untreated wastewater at a WRF is dictated by the sewershed service area. Large sewershed service areas, that are typical of centralized WRFs in many urban areas, can make public health interventions challenging. Sub-sewershed sampling (e.g., from a manhole within the sewer network) or building-scale sampling allows for a more targeted spatio-temporal analysis of SARS-CoV-2 in a community, and thus can help better identify the source of outbreaks and VOCs. A sudden increase in the SARS-CoV-2 viral load or the relative abundance of a particular SARS-CoV-2 lineage can be defined as an outbreak in the context of wastewater monitoring. This is often exactly what is meant when outbreaks are mentioned concerning wastewater analyses. Otherwise, the definition of outbreaks in the classical epidemiological sense is a challenge that cannot be overcome with WWGS alone, even if wastewater analysis can provide very good indications. For example, several universities have implemented SARS-CoV-2 wastewater surveillance monitoring systems to ensure the health and safety of students and faculty. Typically, these sampling locations are established at sewer cutoffs, allowing access to the wastewater leaving campus living facilities (e.g., dorms and campus apartments) or frequently visited facilities (e.g., student unions, libraries, dining areas)[38,42,48]. The indication of locations may allow more targeted analyses and adapted surveillance measures.

### *Virus concentration and RNA extraction methods*

Due to the complexity of wastewater matrices, recovering viral particles can be challenging. Without an effective recovery protocol, downstream quantification may significantly underestimate true SARS-CoV-2 levels. There are several methods to concentrate viral particles from wastewater; however, the most frequently used methods are polyethylene glycol (PEG) precipitation, electronegative membrane filtration, ultrafiltration, and ultracentrifugation[7,39–41,51–53] (Figure 3).

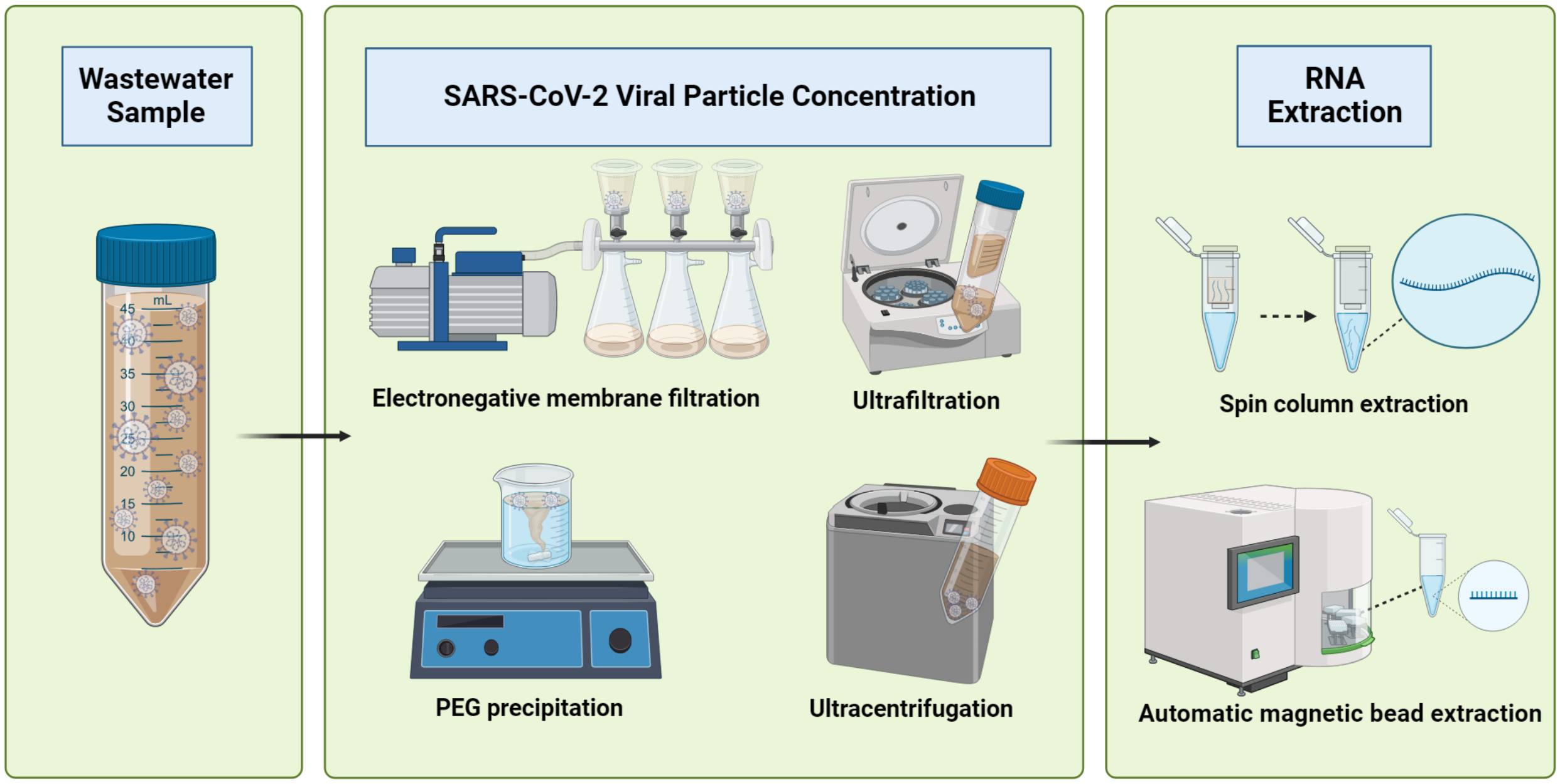

**Figure 3:** The sample collected from Figure 2 is represented as the "Wastewater Sample". Each section describes the most common viral particle concentration and RNA extraction methods employed (Table 1).

PEG precipitation requires the amendment of wastewater samples with a solution of salt and PEG, resulting in a supernatant that contains concentrated SARS-CoV-2 particles. Recovery rates ranging from 46.6 to 62.2% are typical of this method[54–56]. This method provides a reliable and inexpensive option for viral particle concentration, but can be a severe bottleneck in the wastewater analysis workflow. PEG precipitation takes 2 to 6 hours for initial mixing, which requires overnight incubation and a lengthy centrifugation step. A rapid PEG approach, without an overnight incubation step, yields drastically lower recovery efficiencies between 18.8% and 35%[55].

Electronegative membrane filtration in conjunction with a cation conditioning solution (e.g., NaCl or $MgCl_2$) provides a simple, high-speed method to concentrate SARS-CoV-2 viral particles. Typically, the pore diameter of electronegative membranes is between 0.22 and 0.8 µm, thereby accumulating larger particles on the membrane surface. Adding a cation conditioning solution forms salt bridges within the negatively charged membrane, promoting the adsorption of free-floating SARS-CoV-2 virus particles that are significantly smaller than the membrane pore size. This method boasts a high recovery efficiency of SARS-CoV-2, up to 65.7%[45,57].

Ultrafiltration is a direct virus concentration method without conditioning treatment or a lengthy precipitation process. This method differs from electronegative membranes as it concentrates SARS-CoV-2 particles based on size exclusion rather than electrostatic forces, maintaining pore sizes ranging from 5 nm to 0.1 µm down to 3 kDa. While this does seem promising, the viral particle recovery efficiencies are lower than other methods (28-56%)[57]. This method can only process small volumes of wastewater and is prone to clogging. The complexity of wastewater matrices necessitates multiple ultrafiltration units to overcome this, but the equipment and cartridges are expensive and concentrate potential PCR inhibitors alongside SARS-CoV-2 virus particles[57].

Ultracentrifugation is a long-standing method of concentrating viral material by centrifuging the wastewater sample at upwards of 100,000g to create a pellet[57–59]. Although this method provides a quick concentration of viral particles, it co-concentrates inhibitors and relies on larger sample volumes to achieve a large pellet to extract RNA[58]. Further, ultracentrifugation results in consistently low recovery rates of SARS-CoV-2, as low as 19%[57,58].

Following sample concentration, it is necessary to lyse the concentrate via mechanical or chemical methods. Mechanical lysis is typically needed for targets with cell walls. Mechanical lysis is not recommended for virus detection due to the release of nucleic acids from cells, potentially interfering with analyses of viral targets. Chemical lysis through commercially available products (such as Zymo's DNA/RNA Shield) generally suffices for lysing the outer protein coat of viruses, releasing the viral genomic material while reducing interferences from cellular genomic material. Once the samples are lysed, they can be stored, if necessary, without considerable degradation.

After lysis, samples undergo extraction to purify RNA. Several commercially available kits exist, including New England Biolabs Monarch RNA MiniPrep, Qiagen PowerViral DNA/RNA kit, and Zymo Environ Water RNA Kit. The indicated kits yield >70% extraction efficiency when using spiked concentrations of BCoV as a surrogate in wastewater[59]. However, they are column-based extraction kits that require manual extraction, which can increase turn-around time based on the user's experience. Conversely, automated RNA extraction reduces risk of user error and drastically increases throughput. Instruments such as the Maxwell RSC, MagMAX, and KingFisher Flex system offer magnetic bead RNA extraction. Both magnetic bead and column-based extractions have demonstrated equitable numbers of usable sequencing reads[60].

Unfortunately, there is a lack of research investigating different viral concentration methods and sequencing quality. The research that has been conducted is prior to the COVID-19 pandemic, and the methods used in concentrating viral particles are no longer the most frequently used in labs. This knowledge gap can impede future work in WWGS and needs further investigation.

***Quantification methods for wastewater genomic surveillance***

As the COVID-19 pandemic progressed, several molecular tools were employed to quantify SARS-CoV-2. The gold standard of quantification is genomic-based methods, such as RT-qPCR and ddPCR. These methods focus on the gene-specific identification of targets. The detection of target genes is both accurate and highly sensitive, making PCR-based methods a cornerstone of WWGS projects.

RT-qPCR emerged as a powerful tool for wastewater surveillance, allowing for the detection and quantification of SARS-CoV-2. With the addition of a fluorescent dye, a qPCR instrument can measure the fluorescence as the thermal cycler progresses and provide a real-time amplification curve with each cycle. This analysis compares the quantification cycle (Cq) value of a sample with an unknown concentration to a standard curve of known concentrations, allowing for the extrapolation of SARS-CoV-2 virus copy numbers; however, this provides an inherent quantification bias as this method is dependent on the accuracy of the standard curve. Further, due to the complexity of wastewater matrices, amplification and quantification can be affected by inhibitors[61,62].

RT-ddPCR emerged as a strong alternative to RT-qPCR. Instead of comparing to a standard curve, this technique applies Poisson statistics to determine the absolute concentration of the target[63]. Each PCR reaction consists of an oil-water emulsion that partitions each sample into tens of thousands of droplets. Each droplet will either read with a positive or negative fluorescence, and the reader will detect the number of positive droplets. For wastewater, RT-ddPCR has demonstrated a stronger resilience to inhibitors and a higher sensitivity compared to RT-qPCR[61,64–66]. With newer instruments, up to 6 different fluorescent dyes can be detected with ddPCR, enabling an amplitude multiplex of up to 12 targets.

A variation of RT-qPCR developed during the pandemic for detecting SARS-CoV-2 is called Volcano 2nd Generation (V2G)-qPCR[67,68]. The V2G-qPCR method uses a novel polymerase capable of reading both RNA and DNA templates and, therefore, it does not require a separate cDNA synthesis step. Results from V2G-qPCR and RT-qPCR measures are statistically equivalent[69]. Another employed methodology is proteomic quantification detection. Proteomics can provide insight into proteins and their role specific to the target[70]. SARS-CoV-2 RNA genome encodes for at least 29 proteins and can be identified using several types of mass spectrometry analyses[71,72]. While mass spectrometry per sample assay cost may be less expensive, and can provide shorter and cheaper runs than RT-qPCR[71,73,74], RT-qPCR has displayed better sensitivity and specificity[75]. ELISA assays can provide semi-quantitative measurements of specific protein indicators of infection and immunity, such as SARS-CoV-2 specific IgA and IgG[76].

Several primer-probe sets are available to identify SARS-CoV-2, typically in the most conserved regions, such as the N gene[77]. Despite being a relatively conserved region, the N gene is not immune from mutations[78]. As VOC emerges, primer/probe sets become less specific and have a degrading ability to detect positive SARS-CoV-2 samples[14]. Compared to the Index reference sequence from Wuhan strain[79], over 1000 N gene nucleotide mutations have been detected, and more than 300 of them are in commonly used primer sets[80]. Deletions in target genes, such as the N gene in Omicron lineages, can hinder the ability to detect SARS-CoV-2 accurately[78]. Therefore, updating primer sets is an ongoing need to adapt to VOC.

***Advancements in wastewater sequencing technologies***

Sequencing approaches proved effective for detecting mutations and deconvoluting this information to estimate SARS-CoV-2 lineage and sublineage frequencies for WWGS[81–83]. RNA, extracted from wastewater samples, can be reverse-transcribed into complementary DNA (cDNA) and sequenced using various methodological approaches and sequencing technologies[84] to recover as much as possible of the entire viral genome from the wastewater: 1) metagenomics or metatranscriptomics, 2) capture-based sequencing, and 3) amplicon-based sequencing. In metatranscriptomics, the RNA is recovered directly from wastewater samples without any further enrichment for SARS-CoV-2 or depletion of potentially contaminating material from other sources. While metatranscriptomics is a powerful approach for recovering information about whole communities from an environmental sample[85], the downside is that low levels of SARS-CoV-2 are difficult to detect, and much of the sequencing work will go into sequencing other RNA, such as that derived from humans. Previous wastewater metagenomics/-transcriptomics studies showed that genetic material derived from bacteria was more abundant despite additional depletion efforts via size exclusion[85]. As alternatives, capture-based and amplicon-based sequencing, also known as target enrichment approaches, can selectively capture or amplify specific regions of interest from a complex mixture of genetic material[84]. In capture-based sequencing, the target regions of interest (e.g., specific genes or the whole SARS-CoV-2 genome) are selected using capture probes or baits that are complementary to those regions. The capture probes are used to selectively bind and capture the SARS-CoV-2 RNA fragments of interest from a complex wastewater sample. Once the target regions are captured, they can be subjected to library preparation and sequencing. In amplicon-based sequencing, specific regions of interest are selected for amplification and sequencing. Primers are designed to target these specific regions (again, specific genes or the whole SARS-CoV-2 genome), and amplification is carried out using PCR. Such enrichment approaches are particularly useful when the analysis can focus on specific genomic regions or known genes, such as those associated with a particular pathogen like SARS-CoV-2. In the clinical context and genomic surveillance of patient samples, tiled amplicon-based approaches are widely established for sequencing and constructing whole SARS-CoV-2 genomes, e.g., using open-source primer schemes developed and maintained by the ARTIC Network. Since similar protocols and primer schemes can also be used directly for sequencing SARS-CoV-2 from wastewater samples, amplicon sequencing has also become the main approach in WWGS. Amplification generally provides adequate material for sequencing low-abundance viral material out of the wastewater matrix, such as shown via optimized protocols[86] or during low COVID-19 prevalence times[87]. However, amplicon-based methods are vulnerable to primer failure and loss of coverage as new variants arise, and reagents must be continually monitored and updated, similar to reagents used in qPCR and ddPCR assays.

Several sequencing technologies can be used to sequence SARS-CoV-2 RNA from wastewater, each with its own set of advantages and disadvantages. Illumina sequencing is the most widely used sequencing technology for genomic surveillance of SARS-CoV-2 in general[88] and WWGS in particular[89,90]. Short reads produced by Illumina sequencing have a high accuracy, and the platform can generate a large number of reads in a single run. In situations where genomes are reconstructed *de novo,* or large structural variations need to be detected, the major drawback is the limited read length, but this is not so critical when fragmented RNA from wastewater samples is sequenced anyway, and the main purpose is reference-based variant calling. A second short-read technology, more rarely used but also applied in WWGS, is IonTorrent sequencing[81,91,92]. As alternatives, single-molecule real-time sequencing (SMRT) technologies, such as those provided by Oxford Nanopore Technologies (ONT), can produce longer amplicon reads, e.g., approximately 400 bp reads based on an ARTIC Network Protocol[11,93], which can be useful for resolving complex regions of the SARS-CoV-2 genome. In the specific context of WWGS, longer reads can help infer synteny information about mutations that belong to the same viral variant because they are detected on the same read. However, it is challenging to derive long RNA fragments from wastewater samples, and the amplicon approach limits maximum achievable read lengths. Nevertheless, ONT is placed second among the most used sequencing technologies in clinical SARS-CoV-2 genomic surveillance[88] due to its lower initial costs, the putative option to sequence longer amplicons[94], and potential future applications regarding real-time and on-side sequencing. In addition, ONT can also sequence RNA natively without the need for cDNA transcription. SMRT technologies, and in particular ONT sequencing, had higher error rates than other technologies, which may affect accurate variant detection. However, the technologies and thus their accuracy are constantly improving, making them more and more suitable also for accurate variant calling[95].

In addition to technology-related biases, the success of each sequencing technology in recovering most parts of the SARS-CoV-2 genome is highly dependent on the primer scheme used. As with sequencing of patient samples, mutations in the SARS-CoV-2 genome can lead to inefficient primer binding and thus reduced or even absent amplification of the target region, also known as amplicon drop-out. Primer schemes need to be constantly evaluated and adjusted, which is mainly done based on clinical genomic surveillance data. With the decreasing availability of SARS-CoV-2 genomes from clinical genomic surveillance, primer designs may become less accurate and lead to more frequent amplicon drop-outs. In WWGS, due to the mixture of several SARS-CoV-2 variants in the wastewater sample, such failures can be masked very easily and thus go unnoticed An amplicon of a variant that can still be sequenced with the used primer scheme could mask the failure of another amplicon of a different variant that has accumulated one or more mutations in primer sites. Such problems with primer (or bait) designs can be circumvented by metagenomic or metatranscriptomic sequencing, but with the other drawbacks already mentioned.

## Bioinformatics analysis for wastewater sequencing data

In response to the COVID-19 pandemic, the scientific community rapidly developed and 'adapted' bioinformatics tools to address the high amount of SARS-CoV-2 sequencing data, including from WWGS initiatives. The analysis of WWS data, a complex mixture of viral lineages, with low frequencies and fragmented and degraded RNAs, poses unique challenges in the face of bioinformatics methods. While several tools have been developed or adapted for SARS-CoV-2 WWGS (Table 2), many lack validation specifically for WWS data. Therefore, there's a pressing need for rigorous benchmarking of these tools and potentially crafting novel computational methods attuned to wastewater intricacies. Figure 4 outlines a typical workflow diagram used in WWS data analysis.

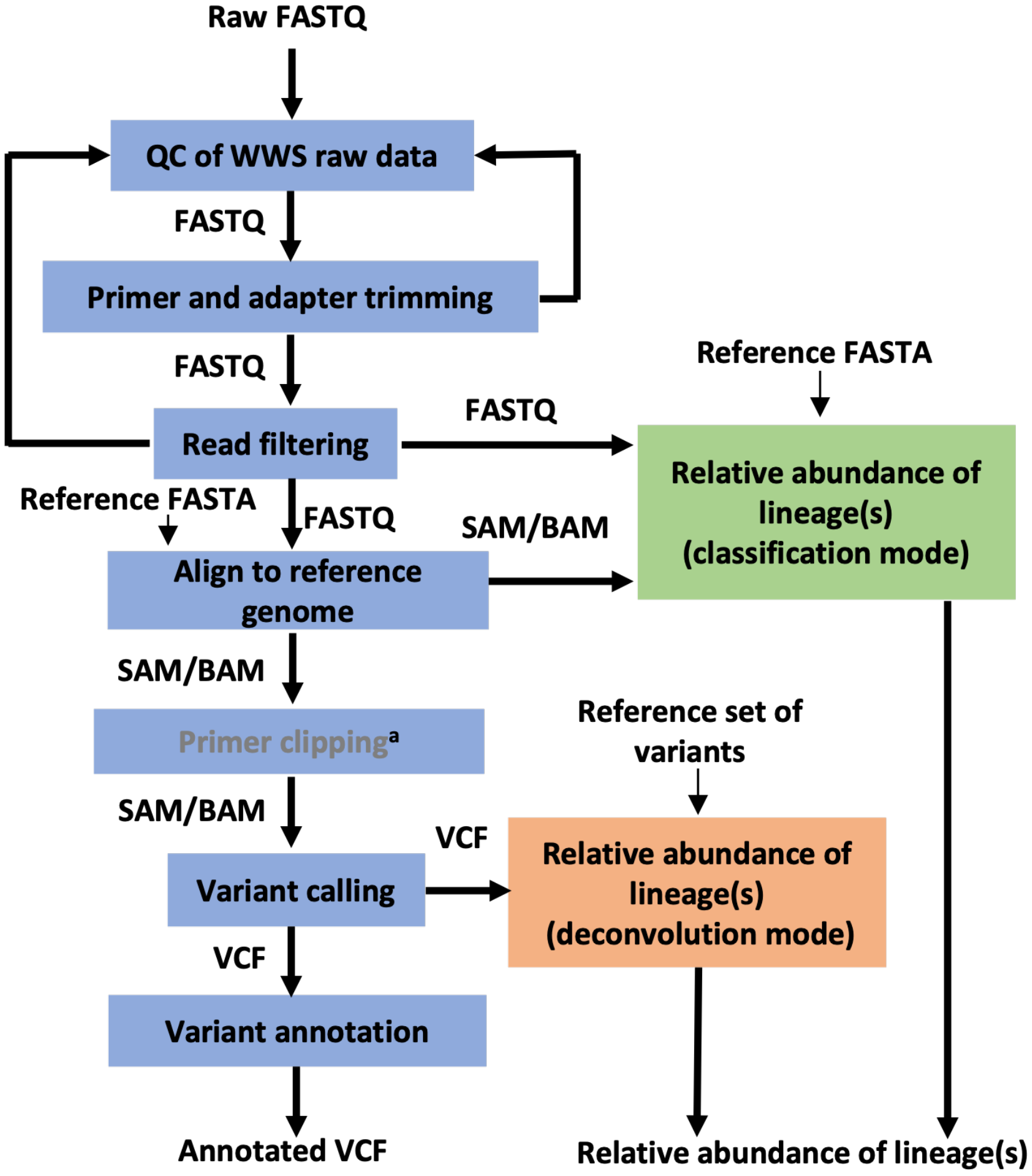

Figure 4. Overview of an exemplary bioinformatics workflow used in wastewater sequencing (WWS) data analysis, ([a]alternative step that can, in comparison to early primer clipping before alignment as shown above, prevent edge effects[96]).

### *Data quality control and error correction tools*

The initial stages of a bioinformatics analysis for WWS data usually include quality control and filtering of the reads, error correction, then trimming of adapters, and subsequently mapping reads to a reference SARS-CoV-2 sequence, primer clipping, and calling mutations (Fig. 4). Conventional error correction tools can be quite challenging when dealing with WWS data reads, as they were primarily optimized for human genome reads and may struggle to handle the subtle variations among viral lineages or sublineages[97]. Several error correction methods tailored for viral sequencing have been proposed to address this issue, such as KEK[98], ET[98], MultiRes[99] or Bayesian probabilistic clustering approach[100].

FastQC[101] and MultiQC[102] are the most popular and efficient tools for quality control in sequencing data analysis and thus also WWS. Ensuring the quality of sequence data is crucial for WWS data analysis, with trimming and filtering being important steps to refine raw reads. For WWS data analysis, tools such as BBDuk[103], Trimmomatic[104], and Trim Galore[105] are largely used for read trimming. iVar[106] is predominantly used for read filtering, while PRINSEQ-lite[107] and fastp[108] are used for managing both trimming and filtering processes based on Phred quality and read length for short reads, and Filtlong[109] for filtering long reads by size. Additionally, viral sequencing data bioinformatics pipelines such as V-pipe[110] or COVID-19 VIral Epidemiology Workflow (C-VIEW)[111] support quality control and filtering steps.

Primer trimming refers to removing primer sequences from raw reads, as they can interfere with downstream analyses. Therefore, trimming them off before performing further bioinformatics analyses is important. Tools like iVar[106], BBDuk[103], cutadapt[112], and Primerclip[113] can perform primer trimming, among other tasks, ensuring that the sequence data is accurate and representative of the native sample. Please note that primer sequences can be removed directly from the raw reads (FASTQ) with the above-mentioned tools or after mapping to a reference genome (SAM/BAM), for example, using BAMClipper[114], see Fig. 4. However, removing the primers after mapping is recommended to avoid errors such as missing deletions near primer sites due to soft-clipping[96]. In addition to this, adapter-contaminated reads can disrupt downstream bioinformatics analysis, including assembly and variant calling. They might induce misalignment or yield incorrect assembly sequences. Therefore, it's imperative to eliminate or trim these adapter sequences from the reads prior to subsequent analysis. Tools commonly used for this purpose, akin to those for primer trimming, include BBDuk[103], cutadapt[112], Trimmomatic[104], fastp[108], and Trim Galore[105].

### *De-novo assembly*

*De-novo* lineage assembly from WWS data for SARS-CoV-2, while similar to viral haplotype reconstruction or quasispecies assembly, faces distinct challenges. Wastewater samples, frequently affected by inhibitory compounds, RNA degradation, and low concentrations, require methodologies that diverge from traditional viral haplotype reconstruction strategies[115]. Due to the ongoing evolution of SARS-CoV-2 and the co-occurrence of multiple closely related (sub)lineages, WWS data often contain sequences with high average nucleotide identity to each other. A robust, tailored approach is essential to address the unique variability in wastewater data quality. Sequencing technologies are capable of capturing this variation, and there are multiple assembly computational methods designed to reconstruct individual genomes from WWS data. Assembly of individual strains from WWS data poses significant challenges, particularly when genetic variation is low, and only a limited number of reads cover variable sites. This often leads to highly fragmented genome reconstructions or consensus assemblies[116,117]. Advancements in sequencing technologies have led to the development of various de novo genome assembly software tools. Prominent among these are assemblers based on the de Brujin Graph approach, such as SOAPdenovo[118], Velvet[119], ABySS[120], IBDA[121], SPAdes[122] and assemblers using the OLC (Overlap Layout Consensus) method, like Edena and VICUNA. Multiple studies have been conducted to evaluate and compare genome assembly benchmarks and performance in the context of viral metagenomics data[123–127]. A notable recent study specifically compared the efficiency of eight commonly used de novo assemblers, utilizing whole genome sequencing data from eight different viruses, including SARS-CoV-2[128]. According to the study, the assemblers SPAdes, ABySS, and IBDA consistently excelled in key metrics such as viral genome recovery, contig number, misassemblies, and mismatches. Another specialized tool is Haploflow, a strain-aware viral genome assembler for short-read sequencing data[129] that was shown to also de novo deconvolute different SARS-CoV-2 lineages from wastewater samples. However, it should also be mentioned that amplicon sequencing data can be particularly difficult to assemble different haplotypes when the overlap between amplicons is small and can lead to confusion between different lineages in a mixed wastewater sample.

***Read mapping***

The read mapping step is fundamental for the accurate identification and subsequent detailed characterization of the viral strains present in the sample. Especially in WWGS, is a critical step in identifying and tracking SARS-CoV-2 (sub)lineages and VOCs. After the preprocessing steps, including read filtering and (optional) primer and adapter removals, the obtained reads are aligned to a reference genome of SARS-CoV-2. This step will enable the identification of viral mutations, via the variant calling tools, and subsequent viral lineage classification. Again, please note that primers can also be removed after alignment to a reference genome to prevent wrong primer clipping and misalignments due to soft clipping. Paired-end reads can also be merged prior to alignment using tools like BBTools[130]. This merging process utilizes the overlapping regions between paired reads for correcting sequencing errors, potentially resulting in sequences of higher quality. Commonly used bioinformatics tools for this purpose include scalable aligners such as BWA-MEM[131], Bowtie[132], and minimap2[133]. Given the dynamic nature of viral genomes and the potential for new variants, regular updates to reference genomes and robust practices are vital to ensure accurate representation and understanding of the SARS-CoV-2 populations in wastewater samples. Incomplete or outdated reference sequence may lead to inaccurate identification of variants, reduced sensitivity to detect new or existing variants, and finally misinterpretation of data and compromised tracking of viral evolution. It is also crucial that the scientific community is relying on a harmonized reference genome for variant calling to share comparable results. At the time of this writing, the community-recognized reference sequence for SARS-CoV-2 is the index consensus sequence obtained from a patient in Wuhan with GenBank accession number MN908947.3 and RefSeq accession number NC_045512.2.

### *SNV calling*

The next step of the pipeline is to identify the SARS-CoV-2 lineages that are believed to be present in the sample and to estimate their relative abundances from a read alignment produced from NGS data of an RNA extract derived from a wastewater sample. In WWS data, the full phasing information of mutations is lost. This is due to fragmentation of the genetic material in the sample, amplification protocols amplifying genomic regions in separate amplicons, and the length of sequencing reads being much shorter than the genome length. In contrast to clinical samples, where we typically assume low diversity and report a consensus sequence representing the dominant inferred lineage, this approach is unsuitable for environmental samples. Specifically, multiple lineages may coexist in wastewater samples, stemming from individuals infected with different lineages. This sample heterogeneity must be considered and is further complicated because these lineages often share mutations, for example, because of parent-child relationships of lineages and sublineages or due to convergent evolution (Figure 5 A). Mutation or variant calling can be performed by a variety of tools also depending on the used sequencing technology, such as iVar[106], SAMtools[134], ShoRAH[135], LoFreq[136], GATK[137], FreeBayes[138], BCFTools[134], Medaka[139], or custom scripts[92]. However, most of these classic variant callers would call variants based on the assumption that all reads belong to the same genome. Mixed-sample variant callers such as ShoRAH expect a limited number of quasispecies which also often have to be estimated before. Thus, accurate variant calling from mixed wastewater populations remains a challenging task. Comparative performance of some of these tools when applied to SARS-CoV-2 wastewater surveillance data has been the subject of published studies[140]. All these variant calling tools have different parameters for filtering according to metrics such as sequencing depth, quality, and allele frequency, impacting the final mutation calls. A recent benchmarking analysis that evaluated variant calling algorithms specifically for WWGS showed that tools like VarScan[141], BCFTools, and Freebayes are generally preferable, particularly when mutations are unknown, due to their higher specificity and sensitivity[140]. However, when specific mutations are known and expected in the output, iVar performed best based on lineage-defining lists of mutations[140].

### *Relative abundance estimation*

Next, the variant calling information (usually in VCF format) can be used to estimate the abundances of SARS-CoV-2 lineages in the wastewater sample. Various computational tools have been developed to accurately estimate the prevalence of SARS-CoV-2 in a population through WWGS. These tools focus on estimating the relative abundance of lineages, based either on a classification approach, such as COJAC[82], VLQ pipeline[142], and expectation maximization EM algorithm for obtaining maximum likelihood estimates of the proportions of different haplotype in a sample[143] or a deconvolution approach, such as LCS[144], VaQuERo[83], Alcov[145], PiGx[146], Freyja[19], LolliPop[147]. The classification approach works at the level of reads and assigns each read (probabilistically or deterministically) to the different reference variants with signature mutations according to the mutations they display. Aggregating the counts of reads assigned to different variants provides an estimation of their relative abundances (Figure 5 B). In contrast, the deconvolution approach takes the individual mutation frequencies computed from the alignment as input. In a mixed sample, the expected proportion of mutated reads at a given locus equals the sum of the relative abundances of variants harboring this particular mutation. Using a reference set of variants, their relative contributions to the observed distribution of mutation frequencies is then estimated by a constrained regression method (Figure 5 C). Some of these methods also allow for considering time dependency in the data by employing different nonparametric smoothing approaches[19,83,147]. Some methods additionally provide confidence intervals for the estimates of variant relative abundances, which is done using bootstrap methods[19,144,147] or closed-form expressions[147].

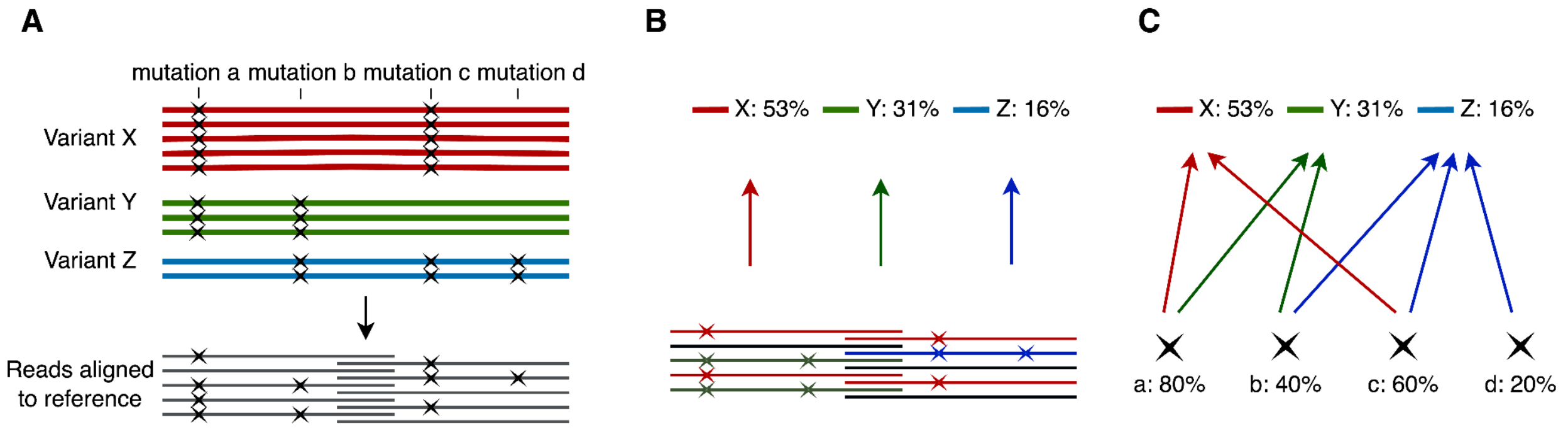

**Figure 5:** Estimating the relative abundances of SARS-CoV-2 genomic variants from wastewater sequencing. **A**: The variants X, Y and Z each have unique but partially overlapping mutation profiles, situated on loci a, b, c, and d. The reads from a wastewater sequencing experiment are aligned to the reference genome, and mutations are called. **B:** In a classification approach, each read is assigned to the variant that most likely generated it. The counts are then aggregated to estimate the relative abundance of variants in the sample. **C:** In the deconvolution approach, the proportions of mutated reads at each variable locus are decomposed into the individual contribution of each variant

***Detection of lineages***

To detect a novel variant or to specify the lineage, haplotype reconstruction methods are used, in which tools classify the mixed read data using different types of methods, including multiple sequence alignment and clustering-based methods, such as ShoRAH[135] and PredictHaplo[148], QuasiRecomb[149] that is based on hidden Markov model, PEHaplo[150] that uses longest common substring, FM-index based search with overlap graph as in case of Savage[151], and rReference-guided assembly used VirGenA[152] tool. All of these methods are reference-based and rely on precise definitions of the variants, which can be generated from clinical sequences generated since the beginning of the pandemic. The reference set can be prepared based on two general approaches to estimate variant frequency - mutation-based (uses a set of marker mutations of lineages) and sequence-based (uses full genome sequence information). It is crucial to acknowledge that both approaches have their limitations. A recent study shows[153] that using sequence-based reference sets can generate higher false positives than the mutation-based approach. On the other hand, the mutation-based approach creates a challenge to select sub lineage defining marker mutations that provide robust assignment in the context of increasing diversity of the lineages. Regardless of the approaches, reference sets of variant genomes may be constructed from existing databases, such as GISAID[154], CoV-Spectrum[155], UShER[156], or NextClade[157]. The selection of appropriate reference datasets is nontrivial, and the results of some deconvolution methods may vary significantly depending on the reference dataset or classification scheme used[153]. Therefore it is important to pay attention to the temporal and geographical range of the references while constructing a reference set[153].

### *Bioinformatics pipelines for WWS data analysis*

Bioinformatics pipelines consist of sequences of automated processes designed to analyze biological data, and they have been extensively used for WW data analysis in different reported studies[11,19,23,146,158]. Predesigned workflows facilitate the complex tasks of organizing, interpreting, and extracting meaningful information from complex WWS data. Some of the most used bioinformatics pipelines include COVID-19 VIral Epidemiology Workflow (C-VIEW)[19], CFSAN Wastewater Analysis Pipeline (C-WAP)[159], PiGx SARS-CoV-2 Wastewater Sequencing Pipeline[146], viralrecon pipeline[160]. Additionally, many researchers resort to custom-developed or modified pipelines tailored to their unique research requirements[161–163]. The widespread adoption of comprehensive bioinformatics pipelines is driven by the necessity to systematically assess a large volume of samples, by ensuring the accuracy and robustness in detecting specific SARS-CoV-2 lineages, determining their prevalence in population, and estimating their relative abundance.

## Applications of wastewater genomic surveillance

When implemented carefully and integrated with other surveillance systems, WWGS can be a highly effective method for understanding and monitoring infectious diseases. By tracking the evolution of viruses, detecting novel mutations, and identifying emerging lineages, WWGS can help identify hot spot clusters and potential outbreaks, especially in smaller to medium population scales. Furthermore, it offers critical insights into community infection rates and can help to assess the impact of vaccination programs, thereby shaping and informing public health strategies and interventions. It is crucial to understand the benefits and limitations of WWGS to make it a robust and powerful public health tool and complement other surveillance systems[164].

### *Outbreak investigation*

An outbreak can be defined by the occurrence of more cases of a disease than expected in a given location and over a given period of time[165]. In the context of SARS-CoV-2, an important application of bioinformatics analytics is the study of transmissions of the virus and its evolutionary relationships with existing VOCs, for example, to detect outbreaks. While WWS data can show increases in SARS-CoV-2 mutations or known lineages at a given location over time, they lack the link to epidemiologic data, so they cannot be used alone to detect outbreaks in the classical sense. In addition, wastewater samples for genomic surveillance can be collected and analyzed at different population scales, ranging from small (like a retirement home) to medium (like a small town) to large (like a county). While WWGS from small to medium-sized settings can help in the early detection of isolated outbreaks, additional surveillance systems need to be integrated, especially in more complex settings and population structures. While clustering and phylogenetic methods exist to detect transmission and characterize evolutionary relationships, most are tailored to the analysis of complete genomes derived from sequencing data of diagnostic samples. In addition, genomics data from diagnostic samples can be integrated with epidemiological data, which is crucial to detect and clearly define outbreaks[166]. Then, phylogenetic analysis of genomic sequences can calculate the distance across the closest pairs to trace the evolutionary lineage of viruses.

A directed network of the viral outbreak is required to assess the direction of transmission and detection for a superspreader event. This is pivotal to finding transmission clusters and limiting them to the containment zone. The transmission clusters are also affected by the presence of multiple genomic variants of a virus within an infected individual, referred to as viral quasispecies. Early transmission inference methods primarily did not account for intra-host viral diversity. However, accounting for intra-host diversity particularly facilitates the identification of viral evolution directionality in instances where reliable phylogenetic rooting is challenging, which is common in epidemics such as the SARS-CoV-2 outbreak involving multiple viral introductions. This was addressed later in phylogenetic approaches for inferring transmission direction tools such as Phyloscanner[167], SeqTrack[168], TNeT[169], TiTUS[170], SharpTNI[171], VOICE[172], and BadTrIP[173]. Some tools took a step further and integrated models of social/contact networks to consider the social aspects of outbreaks, such as QUENTIN[174] and SOPHIE[175].

Nevertheless, most existing models are not applicable when it comes to WWGS, since the detection of transmission networks is not feasible solely based on WWS data. Particularly concerning the nature of WWS data, where sequences lack host labels and the reconstruction of robust consensus genomes is challenging, it is imperative to reevaluate the incorporation of both inter- and intra-host diversities.

On the other hand, analysis and forecasting of evolutionary dynamics of SARS-CoV-2 have been carried out using various methods, from phylogenetics to machine learning[176–184]. Previous models designed to predict and monitor the dynamics of SARS-CoV-2 using WWS data typically incorporate additional epidemiological data, such as case numbers and hospital admissions specific to the particular locations. When relying solely on WWS data, these models primarily forecast epidemiological measurements like hospital occupancies or the number of cases within a community. While there has been extensive research into variant and mutation detection and forecasting using clinical SARS-CoV-2 data, such exploration of wastewater data remains relatively limited in current studies and methods. Many existing studies on clinical SARS-CoV-2 data use phenomenological growth models for transmission analysis and forecasting SARS-CoV-2 trajectories, for instance PyR0[179] model, which is a hierarchical Bayesian multinomial logistic regression model was designed for the detection and analysis of emerging SARS-CoV-2 variants with enhanced fitness. Nonetheless, this model as well as most of such forecasting models primarily concentrates on predicting the emergence of individual mutations. While the non-additive phenotypic effects of combinations of SARS-CoV-2 mutations have been suggested to be responsible for the non-linearity of SARS-CoV-2 evolution that significantly complicates its dynamics and, therefore, its forecasting[185–187]. HELEN[188] tried to address this by examining the community structure of viral coordinated substitution networks of the SARS-CoV-2 as a case study. This computational framework identifies and merges densely connected communities of single amino acid variation (SAV) alleles into haplotypes. It utilizes statistical inference, population genetics, and discrete optimization techniques, which could efficiently predict emerging VOCs and VOIs months before their prevalence reaches noticeable levels. Extending such models to be applied to wastewater data would hold significant importance as a supplementary resource to more traditional surveillance methods, contributing to more comprehensive public health monitoring efforts and gaining insights into the virus's transmission patterns and evolutionary trajectory prior to the availability of clinical data during an outbreak.

### *Monitoring of viral evolution and detecting novel mutations*

WWGS offers an additional, independent, non-invasive resource for tracking SARS-CoV-2 evolution, which is crucial for long-term adaptation to co-existence with this pathogen and its continuous control to decrease the COVID-19 health burden in the post-acute pandemic period[189]. This is of particular value during the phase of reduced clinical surveillance, lifted restrictions, and increasing genomic diversity, with different viral sublineages in side-by-side circulation and higher odds for co-infections and recombination events[190]. WWGS can detect the emergence or introduction of known (sub)lineages in particular regions weeks prior to their identification in clinical samples, subsequent monitoring of their contribution to SARS-CoV-2 infections at the population level, prediction of the reproductive advantage, and further accumulation of novel mutations and putative novel so-called cryptic lineages[19,90,158]. This allows viral trees that have evolved over time and among various regions to be recognized and compared. Identifying novel mutation signals and potential (sub)variants through WWGS may even prompt their increased and targeted clinical surveillance[191], indicating that both approaches are complementary and can strengthen the viral monitoring network. However, WWGS is likely superior in regions with limited genomic surveillance of SARS-CoV-2 due to non-challenging sample collection and the ability to generalize data for a particular area without a need for mass sample sequencing, contrary to clinical surveillance[192].

Earlier characterization of amino acid substitutions in spike protein and other viral proteins through WWGS offers a more swift initiation of experimental studies on immune escape mutations and drug resistance, pivotal in vaccine-adaptation efforts and predicting the efficiency of authorized direct-acting antivirals. It also enables the initiation of *in vivo* research on the clinical relevance of novel (sub)variants and particular mutations, which is of utmost importance considering that the intrinsic severity of future SARS-CoV-2 variants remains uncertain[193]. Using WWGS to detect more severe viral variants, e.g., harboring mutations enhancing fusogenicity, would allow for more targeted and rapid public health responses, translating into decreased morbidity and mortality. Furthermore, WWGS could be employed to track mutational signatures from exposure to mutagenic antivirals (i.e., molnupiravir authorized in selected world regions)[194], essential to explore the impacts of such treatments on the trajectory of (sub)variant generation and onward transmission. Moreover, WWGS is a tool to track the cryptic circulation of SARS-CoV-2 variants that may appear entirely deescalated using clinical surveillance but may otherwise re-emerge or lead to the generation of new lineage, e.g., through recombination events[195]. Lastly, WWGS can support the early detection of spillback of mutated variants that could arise during viral circulation in non-human reservoirs that SARS-CoV-2 has already established (e.g., free-ranging white-tailed deer[196]). The clinical consequences of such retransmission to the human population are challenging to predict since mutation-driven adaptations to a new host may lead to decreased adaptation to the human environment but also to improved evasion of acquired immunity, including cellular response, and thus higher susceptibility to severe disease[197,198]. Therefore, detecting such events as soon as possible can guide other surveillance systems and is necessary to implement effective containment measures.

As SARS-CoV-2 is far from eradication and continues to evolve, while the risk of the emergence of novel, clinically relevant viral variants remain high, implementing WWGS to detect them ahead of their effective spread in the community is essential. Although WWGS is increasingly applied in this regard, the results are primarily made available through peer-reviewed literature. Ultimately, WWGS should serve as an early warning indicator for detecting and emergence of novel mutations and associated sublineages, functioning in tandem with other surveillance systems. However, considering its value and ongoing transition from the acute phase of the COVID-19 pandemic, it is pivotal to establish a global public repository of SARS-CoV-2 sequences generated with WWGS over time in various world regions, enabling genomic epidemiology and real-time surveillance to monitor the emergence and spread of viral sublineages in a fashion similar to GISAID. This would increase the relevance of WWGS to global COVID-19 research and guidance of public health measures and policy, including recommendations on maintaining or updating COVID-19 vaccine composition for primary or booster doses.

***Tracing emerging lineages and monitoring infection rates***

Wastewater-based epidemiology can be used as an early warning tool for viral spread in the community, identifying emerging lineages, and monitoring infection trends, potentially informing public health actions and policy decisions. Contrary to clinical surveillance, it is not biased toward symptomatic infections and is not affected by individual engagement in testing. Instead, it can be applied to estimate the temporal and spatial trends of total (including undiagnosed) infection load at the community level[199]. Since the beginning of the COVID-19 pandemic, wastewater epidemiology, particularly based on quantitative assessment of genomic copies, has been applied to detect SARS-CoV-2 for community-wide surveillance and in smaller catchments for more targeted surveillance[200–204]. The role of such an approach in routine monitoring of infection trends, emerging lineage detection, and cluster identification is even increasing during the transition from the acute phase of the pandemic when clinical surveillance is no longer as extensive, restrictions are lifted, and the public is generally less concerned about the COVID-19 threat. While routine WWGS can never fully replace other surveillance instruments, it can, under such conditions, become a primary source of information on trends of SARS-CoV-2 circulation in various communities. A recent study from the USA showed that forecasting models derived from wastewater-based epidemiology data can accurately predict the weekly new hospital admissions due to COVID-19, providing a 1-4 weeks window for introducing mitigation measures[205].

The qualitative assessment offers additional advantages in this regard. Tracking the dynamics of the contribution of particular (sub)lineages in wastewater is a powerful early warning tool to understand viral shifts that occur at the community level on different scales. Their spatial and temporal spread can be tracked, in real-time or retrospectively, by integrating data derived from various catchment areas, allowing for the identification of hot spots of specific viral (sub)lineages[162,206,207]. Foremost, qualitative WWGS can detect them much earlier than clinical testing, ahead by weeks or even months[19,208–211], enabling expedition of an effective containment response and by guiding, as a complementary surveillance strategy, public health policies regarding face masking, booster vaccinations, and/or decreased social mobility. Ultimately, WWGS, coupled with (sub)lineage-oriented risk assessments, can become an effective and complementary tool to decrease infection rates, long-term consequences of COVID-19, hospital admissions, and mortality.

Moreover, WWGS has the potential to screen cross-border SARS-CoV-2 spread. Applied to aircraft wastewater samples, it can effectively monitor viral (sub)lineages potentially carried by onboard passengers and enrich data on viral diversity in departure areas. In the past, selected SARS-CoV-2 variants were detected in clinical samples from returning overseas travelers[91,212]. Therefore, establishing a global aircraft-based WWGS network is postulated with use in the context of COVID-19 and future viral threats[213]. Such a network could compensate for limited genomic surveillance in various world regions, particularly low- and middle-income countries, which is essential to counter the threat of future viral variants[192].

***Assessing the effectiveness of vaccinations***

In the post-acute pandemic era, COVID-19 vaccination remains an essential and primary public health intervention to decrease SARS-CoV-2 morbidity and mortality. Omicron sublineages are clinically milder, but their infections can lead to severe outcomes in selected patient groups, causing health and economic burdens, management of which requires appropriate preparedness[214,215]. However, vaccine-induced humoral immunity is short-lived, while the virus accumulates immune escape mutations, justifying booster dose recommendations and vaccine updates. At least one booster dose will likely be recommended annually, particularly for the elderly, patients with comorbidities and immune deficiencies, and healthcare workers[216].

Wastewater-based epidemiology can be used to evaluate the efficacy of vaccination by showing a decrease in SARS-CoV-2 RNA positivity in response to immunization, as was successfully demonstrated in the initial phase of mass COVID-19 vaccination[217,218]. However, as with all analyses of wastewater data, careful consideration must be given to other potential confounding factors that may affect RNA concentrations in addition to vaccination. Nevertheless, the possibilities arising from such analyses have yet to be fully exploited in the context of vaccination. Similar studies following subsequent booster administration, integrating data on vaccination coverage in particular areas, could reinforce confidence in COVID-19 vaccinations, especially when resources for real-time tracking of vaccine effectiveness are available to the public in limited form. Such an approach could also be employed in specific settings, e.g., hospitals or nursing homes, before, during, and after booster vaccination campaigns, enabling a better understanding of the effect of immunization on virus spread in the community. WWGS provides further opportunities, as it can offer to track the effect of vaccination on particular sublineages, which are in concurrent circulation but may differ in sensitivity to neutralization antibodies elicited by vaccines as observed currently within the Omicron lineage[219,220]. Using WWGS, such data could be obtained earlier than through clinical surveillance and epidemiological analyses but, again, needs to be interpreted carefully and in light of the population scales of wastewater sampling. However, this can be of particular use if one considers that even with an mRNA platform, the time needed to develop and authorize an updated vaccine may be enough for SARS-CoV-2 to generate progenitors that diverge from the selected antigen, causing public concern over the vaccine's effectiveness. Therefore, WWGS may be the first to provide an initial assessment of its performance on the population level, which may be valuable in decreasing vaccine hesitancy. Furthermore, WWGS can provide a more accurate assessment of vaccine effectiveness on the population level than analyses based on cases of breakthrough infections with presenting clinical symptoms. Lastly, since SARS-CoV-2 eradication is highly unlikely with currently available vaccines, WWGS could generate data on which mutational sites are positively selected under increased immunization levels due to booster administration.

In addition, data generated through WWGS can be integrated into the system of continued monitoring of the evolution of SARS-CoV-2, which is pivotal in guiding antigen selection for updated COVID-19 vaccines. Of note, none of the authorized COVID-19 vaccines is based on attenuated live SARS-CoV-2; thus, shedding of the vaccine-derived virus will not confound WWGS with false positive signals[221], although such a possibility needs to be considered if replication-competent vaccines would become available

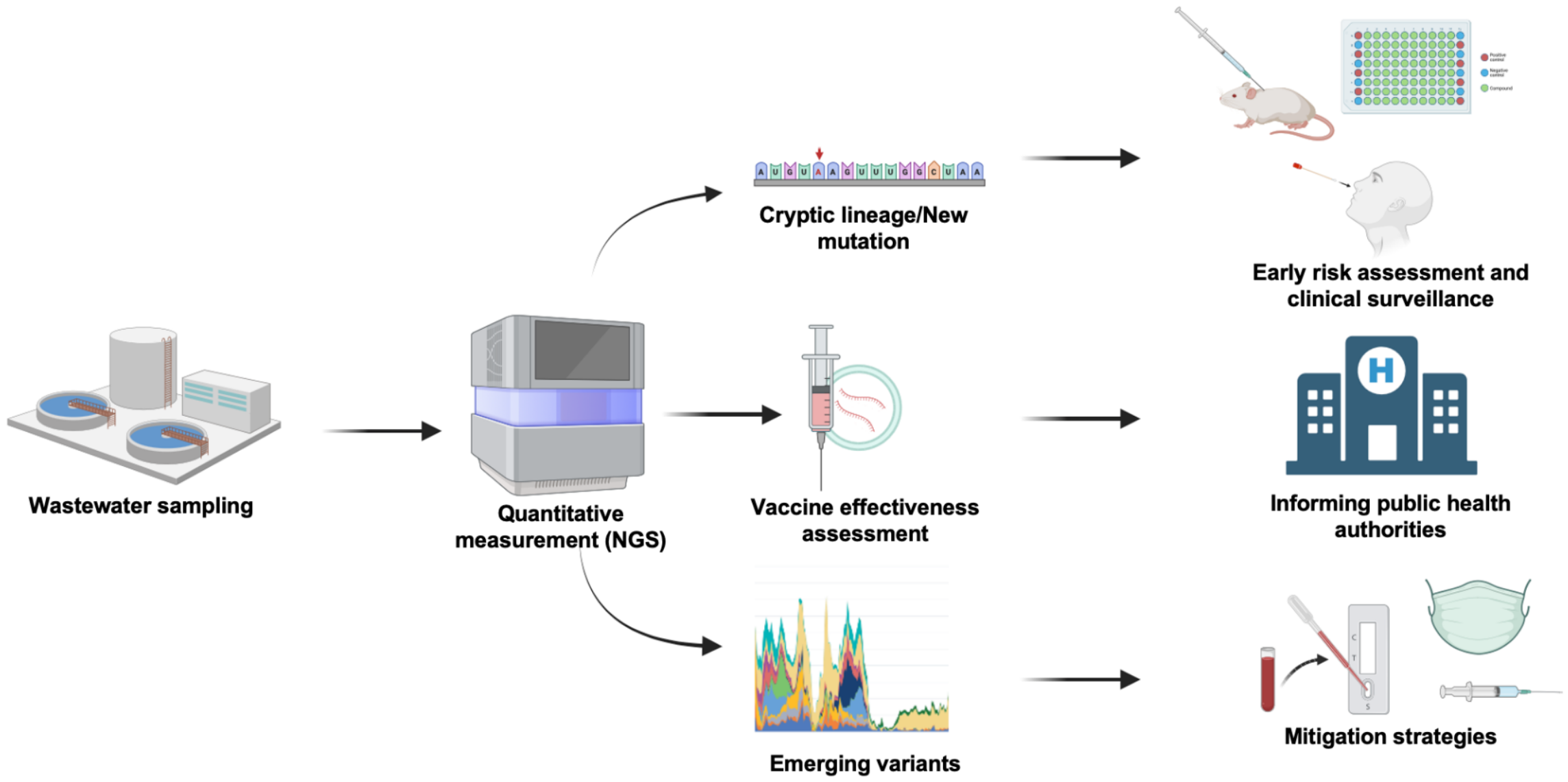

**Figure 6.** Main applications of wastewater genomic surveillance and their potential impacts on risk assessment, public health guidance, and mitigation strategies. New mutations discovered with the help of WWGS can serve as a basis for virological experiments and clinical genomic monitoring, e.g., by adapting variant-specific PCR designs. WWGS can also be used to study vaccine efficacy. However, several confounding factors must be considered, such as other environmental factors and the changing immune landscape of the sampled population. WWGS can also serve as an early warning system by continuously monitoring SARS-CoV-2 lineages and detecting increasing trends, especially in the face of reduced clinical testing and genomic monitoring of patient samples. However, it is important to emphasize that WWGS can inform public health authorities on these health aspects. However, due to its still academic nature in many respects, it cannot be used alone to inform disease containment strategies and policy decisions. Further research in the field of WBE and WWGS is needed and translation into robust public health surveillance systems.

## Challenges in wastewater genomic surveillance

The analysis of WWS data is more complicated than clinical samples collected through nasopharyngeal swabs. Discrepancies between clinical and WW samples may arise from several factors. Firstly, the mixed strain composition in WW samples complicates the analysis. Secondly, the viral genomes in these samples tend to be more degraded, often hindering the ability to obtain a consensus genome sequence for each sample. Additionally, the lower virus titer found in WW samples can impact the effectiveness and accuracy of bioinformatics methods. The complex nature of wastewater is also reflected in the WWS data quality.

A common challenge in WWGS is amplicon dropout, where certain genomic regions are underrepresented or under-sequenced. These under-sequenced regions (USRs) result from mismatching primers resulting from viral mutations and are identified by the presence of unknown bases (represented by 'N') in consensus sequences. Amplicon dropouts and coverage bias have been reported with the sequencing protocols based on ARCTIC[222], as well as Midnight[223] panel primers. Mutations within the primer binding site can prevent primer-annealing, leading to dropout or loss of that amplicon. Additionally, primer-primer interactions could result in amplification bias of interacting amplicons[222–224], resulting in coverage bias and consequently affecting the identification of mutations in the viral genome. This can result in incomplete information, hinder accurate lineage nomenclature and classification, and lead to inaccurate assessments of viral diversity. Variant callers exhibit a 'blind spot' for USRs, potentially leading to the omission of critical variants in wastewater samples. Consequently, computational methods for quantification that rely on a deconvolution approach may yield suboptimal performance. Additionally, classification-based approaches risk generating higher false positives by assigning the reads to non-specific lineages. As the virus's genomes evolve, as a general recommendation, it will likely be necessary to update and adjust the primer panels to maintain optimal coverage for all genomic regions. Emerging lineages must be monitored for disruptive mutations so that primer schemes can be updated routinely. Spike-in primers can be effective stopgaps to maintain performance between major designs. One effective approach to facilitate screening a large number of genome sequences and identify relevant mutations is to use the PCR-strainer pipeline[225].

In computational methods for estimating relative abundances, the number of reads that map to a particular (sub)lineage reference or the number of reads containing a specific variant associated with certain (sub)lineages are used as indicators of the relative frequency of these (sub)lineages within the sample. However, there are two main reasons why this assumption is not always correct. First, most library preparation methods include polymerase chain reaction (PCR), which is known to amplify some fragments more efficiently than others, depending on their GC content and length[226,227]. Second, in Illumina technologies, there is a specific type of duplicate reads called optical duplicates that arise from miscalling a single cluster as two separate clusters, or there is a probability that one molecule of the library initiates two independent clusters[227]. These factors can lead to biases in the number of reads, resulting in an inaccurate representation of the initial abundance of RNA molecules. These biases are particularly critical for the highest and the lowest abundant RNA types. As of the writing of this paper, no studies have been found that investigate the impact of deduplication methods on the relative quantification of (sub)lineages in wastewater samples. However, research in the field of RNA-seq quantification for gene expression has revealed that for single-end reads, using molecular indices for deduplication is essential for accurately identifying differentially expressed genes. Furthermore, the usage of pair-end reads and deduplication based on two indices can markedly increase the sensitivity and accuracy of quantification analysis[227]. Future benchmarking projects should prioritize addressing challenges like amplicon dropout and optimizing read deduplication strategies in bioinformatics methods. These areas are essential for enhancing the accuracy and reliability of WWGS for SARS-CoV-2.

The robustness of SNV-based and sequence-based methods for bioinformatic analysis of WWS data heavily relies on the composition of the wastewater sample. Sequence similarities among related (sub)lineages can cause ambiguity in variant detection. Thus, the set of reference data, i.e., the considered lineages and selected characteristic mutations/sequences that the WWS data are compared against, impacts which lineages and sub-lineages can be identified and how specific variant calls/reads can be assigned. Current bioinformatic methods already implement various approaches for reference reconstruction. VLQ[142] selects reference lineages based on the spatio-temporal context of the wastewater sample and samples a specific number of genomic sequences for every lineage according to a predefined threshold for the genomic variation that should be captured[142]. Freyja reconstructs a set of characteristic lineage mutations based on the UShER phylogenetic tree[19], while other SNV-based tools like wastewaterSPAdes and SAMRefiner rely on a rule-based selection of characteristic sets of mutations considering lineage-differentiating power[228,229]. However, the reference bias remains strong and requires continuous awareness and manual review[153]. Because of the fast evolutionary changes of the virus, reference data need to be re-evaluated for every sample and pandemic timeframe. Specifically, convergent evolution and novel lineages challenge the current strategies for reference reconstruction: depending on the circulating lineages of interest, it becomes more challenging to represent genomic variation and still guarantee sufficient differentiation power between sub-lineages. Furthermore, most currently applied tools rely on a large amount of clinical sequence data to reconstruct their reference data sets. Decreased clinical sampling poses a challenge for bioinformatic WWGS and should be considered for further research in method development, especially in terms of identifying and quantifying unknown lineages.

Early identification of unknown viral variants based on novel genomic signals represents one desired benefit and also a great challenge for WWGS. Currently, novel variant detection is mostly conducted retrospectively, while real-time cryptic variant detection represents an ongoing bioinformatics research topic where the first approaches are slowly published. Previously, CryKey was developed as one of the first tools for non-retrospective cryptic variant detection[230]. CryKey identifies cryptic variants based on sets of mutations that co-occur on the same reads but have not been observed to co-occur before in clinical sequence data. The tool addresses bias and artifacts in WSD by rule-based filtering of mutations and reconstructs a reference table mapping SNP information and lineage assignments from clinical sequence data. Overall, biases of WWS data and their epidemic context should be continuously monitored and considered during bioinformatic methods development.

## Ethics and privacy concerns

Using the WWGS technology could significantly improve our ability to detect, monitor, quantify, and measure the spread of SARS-CoV-2. However, the WGS technology used for public health monitoring raises various legal and ethical concerns. Initiatives by state or local health authorities to monitor wastewater for SARS-CoV-2 would likely fall under the state's well-established police powers. These powers typically grant state and local governments extensive latitude to enact and regulate measures to safeguard public health and welfare.

While traditional public health surveillance primarily targets identifying individual cases via testing, screening, reporting, and contact tracing, wastewater surveillance also likely falls under these expansive state powers. Yet, there are inherent privacy issues linked with wastewater surveillance. In the US, wastewater surveillance presents inherent privacy concerns, potentially challenging the Fourth Amendment's protections against unreasonable searches and seizures. Historically, courts have ruled that there's no reasonable privacy expectation in wastewater once it enters the public sewer system[231–233]. However, these decisions haven't ruled out every expectation, especially when individual rather than corporate interests are considered[232]. Given recent judicial perspectives on the Fourth Amendment in relation to involuntarily shared, yet intimately personal data, certain wastewater collection practices might be deemed as searches. As the data derived from wastewater becomes increasingly detailed, reaching levels as specific as individual households, its surveillance might intersect with Fourth Amendment concerns[234]. The WWGS programs, be they individual tracking, location monitoring, or the use of aggregate data, are likely to fall under the special needs doctrine because they aim to address public health challenges rather than to effect the goals of traditional law enforcement[235]. Its constitutionality becomes more likely if it provides a broad overview without pinpointing specific individuals. Finding the right balance between individual privacy and overarching public health objectives is crucial. WWWGS touches on ethical dimensions like personal autonomy and beneficence. While individuals generally have the right to manage access to their health information, this must be weighed against the benefits of quicker disease detection at the population level via wastewater screening[236].

Utilizing the data acquired from wastewater sampling for public health objectives might introduce further legal and ethical issues, extending beyond the simple collection and observation of the data[237]. Public health authorities could decide to use evidence of the presence of SARS-CoV-2 in wastewater to recommend multiple intervention strategies, such as tracking the spread of increasing lineage clusters and identifying previously unrecognized areas to enable better targeting of resources and supportive infrastructure for communities affected by COVID-19. Further, based on such information, public health authorities could recommend increased testing among people living in such areas or implementing neighborhood-wide resources to provide voluntary screening programs. In addition, voluntary screening programs facilitate individuals' autonomy and right to refuse testing. On the other hand, a conditional screening program requiring individuals under a quarantine order to undergo testing before leaving their homes would require a delicate balance between the state's exercise of police powers and an individual's right to decline testing. Established legal precedents indicate that mandates for testing or treatment related to highly contagious diseases are typically within the jurisdiction of a state's legal authority[238]. If individuals opt not to comply, they wouldn't be physically compelled to undergo testing. Instead, they could face alternative measures like extended quarantine or reasonable financial penalties based on the policies and regulations of the respective country. In general, country-specific public health and policy guidelines are needed to protect privacy and personal rights and maximize reasonable opportunities to use wastewater analysis to improve people's health.

## Conclusions

Genomic sequencing of wastewater samples has emerged as an effective tool for infectious disease monitoring due to key aspects such as community-wide sampling, early detection of emerging threats, cost-effective surveillance, its non-invasive and unobtrusive nature, and complementing other well-established surveillance systems such as syndromic surveillance and genomic sequencing of clinical samples. Optimized sequencing protocols and tailored bioinformatics methods to address wastewater-specific genomic data should be developed to make this feasible. Many tools have already been developed for similar problems in genomics, but it is imperative to perform comprehensive benchmarking before they can be applied to genome-based wastewater sequencing. In addition, and especially during the COVID-19 pandemic, customized tools for analyzing wastewater sequencing data of SARS-CoV-2 were rapidly developed and implemented. Continuous benchmarking will allow not only an understanding of the quality of state-of-the-art methods but also help to determine the future direction for methods development and their current feasibility.

Genome-based wastewater surveillance is an excellent supplement to clinical or epidemiological monitoring of pathogens' spread. However, it is not mainstream yet. Currently, only 70 out of 194 countries use wastewater surveillance[239]. Developing countries do not have the resources to sequence several samples of the population to trace emerging variants of SARS-CoV-2[240]. An appealing alternative to that can be collecting and sequencing viral samples from wastewater, which is significantly more cost-effective and expands the coverage of a surveilled population.

A typical COVID-19 wastewater surveillance program is a powerful epidemiological tool that provides quantification of SARS-CoV-2 and acts as an early warning system for community infections[183,241,242]. While wastewater genomic surveillance can never fully replace other typical surveillance systems, it provides similar or complementary assurances while also generating sequencing data, which can be used for novel mutation or cryptic VOC detection. To make it more cost-effective, pooled sequencing and advanced algorithmic processing can be used. Pooling will increase the number of samples sequenced in a single run. It should be noted that computational methods for inference of heterogeneous viral populations from pooling data exist[243,244], but should be benchmarked and adjusted to the specifics of wastewater surveillance data. Novel bioinformatics pipelines specific to wastewater surveillance can be developed to detect potentially novel lineages and their abundances. Currently, universal guidelines are not established to collect wastewater samples, concentrate viral particles, extract RNA, and quantify viral loads. Standard operating procedures (SOP) should be defined, and data can be shared on public repositories just like clinical data repositories[245]. That data can further help us detect putative novel variants before they appear in a large population, and preventive measures can be taken. Wastewater data can help identify the relative abundance of existing VOC and potentially assemble novel mutation profiles, hinting toward emerging lineages. Additionally, wastewater can be used to monitor other viruses without a significant increase in the cost of monitoring, including Influenza A and B, mpox, and norovirus[246–255]. Another notable instance is the surveillance of polioviruses, which historically pioneered the use of wastewater monitoring prior to the SARS-CoV-2 pandemic. However, careful benchmarking and adjustments of sequencing protocols and bioinformatic pipelines is necessary. Nevertheless, all the current initiatives exploring wastewater-based surveillance possibilities indicate its tremendous potential for reliable viral surveillance. Wastewater-based genomic surveillance can be a powerful supplement or even a main methodology for cost-efficient and reliable surveillance of current and future viral pandemics.

**Table 1:** Collection of laboratory methods used for SARS-CoV-2 wastewater-based genomic surveillance (WWGS).

| Method | Available options |
|---|---|
| **Sampling Sources and Scales** | WRF: Raw Sewage[11,37,82,83,90,162,206,211,256–264]<br>WRF and Hospital: Raw Sewage[265]<br>Manholes: Raw Sewage[266]<br>WRF and Manholes: Raw Sewage[267]<br>WRF: Sludge[142]<br>WRF: Raw Sewage and Treated Effluent[89]<br>Building and WRF: Raw Sewage[19,142,268] |
| **Samling Type and Frequency** | Grab[90,142,206,269]<br>Grab: 1x, 2x a week[211]<br>3-hour composite[267]<br>24-hour composite[11,37,83,162,256,263–267,267,269,270]<br>24-hour composite: 1x[258,261,262], 2x[257,259], 3x[19,267], 5x[19] a week<br>Grab and 24-hour composite[82,89,268] |
| **Concentration Technique** | Ultrafiltration[11,82,256–259,265,266]<br>Ultracentrifugation[206,267]<br>PEG precipitation[83,162,211,260,261,267]<br>Electronegative Membrane Filtration[262,263,268,270]<br>Centrifugal Ultrafiltration[37,89,264,269]<br>AI(OH)3 precipitation[89]<br>Affinity Capture Magnetic Hydrogel[89] |
| **Extraction Method** | QIAamp Viral RNA mini kit[82,211,260,264,267,269]<br>RNeasy mini kit[11,265,266]<br>NucliSens Kit[206,256,258]<br>RNeasy PowerMicrobiome Kit[257]<br>MagMAX-96 Viral RNA isolation kit[259]<br>Viral RNA/DNA Concentration and Extraction Kit from Wastewater (Promega)[83]<br>Wizard® Enviro Total Nucleic Acid Kit (Promega)[271]<br>Direct-zol 96 MagBead RNA kit[83,261]<br>Chemagic Prime Viral DNA/RNA 300 Kit H96[262]<br>Qiagen AllPrep DNA/RNA minikit[270]<br>NucliSENS easyMAG[263]<br>EZ1 Virus Mini Kit[37]<br>NucliSENS MiniMag Nucleic Acid Purification System[89]<br>Maxwell RSC Pure Food GMO and authentication kit[90]<br>Zymo QuickRNA-Viral Kit[268]<br>Magnetic-bead based with MagMAX Viral/Pathogen II Nucleic Acid Isolation Kit[268]<br>Zymo Quick-RNA Fecal/Soil Microbe Microprep Kit[142] |
| **Quantification Method** | RT-qPCR[11,19,37,83,89,90,162,206,211,256–261,263–268,270]<br>ddPCR[83,142,257,258]<br>dPCR[266]<br>V2GqPCR[268] |
| **Sequencing Panel** | ARTIC v1[264], v3[11,37,83,89,90,142,211,259,262,263,268,272], v4[266,267] and v4.1[266] panel<br>Swift Nomalase Amplicon SARS CoV-2 Panel (SNAP)[19,162,261,265]<br>Tiled amplicon approach[256,257]<br>EasySeq™ RC-PCR SARS CoV-2 WGS kit v3.0[206]<br>VarSkip 1a panel[83]<br>NEBNext Fast DNA Library Prep Set for Ion Torrent and Ion Xpress Barcode Adapters[260]<br>NEBNext VarSkip Short SARS-CoV-2 Primer[262]<br>Illumina Respiratory Virus Oligo Panel[270]<br>CleanPlex SARS-CoV-2 FLEX Pane[269] |
| **Sequencing platform** | Illumina[19,37,83,89,90,142,162,206,211,256,259,261–266,268–270,272]<br>Oxford Nanopore[11,256,257]<br>Ion Torrent[92,260]<br>Sanger[11] |

**Table 2:** Collection of bioinformatics methods used for SARS-CoV-2 wastewater sequencing data (WWS data) analysis. This overview comprises tools that were developed for the general analysis of sequencing data, adapted for WWS data, and specifically developed for WWS. The overview is not exhaustive but provides a starting point for most representative and popular bioinformatics tools for WWS data analytics.

| | **Software** | **Source** |
|---|---|---|
| **Data QC, primmer trimming trimming and adapter removal, paired-end reads merging** | BBMap (BBDuk)<br>FastQC<br>MultiQC<br>iVAR<br>fastp<br>PRINSEQ-lite<br>Trim Galore<br>Trimmomatic<br>cutadapt<br>Primerclip<br>BBTools<br>BAMClipper | https://github.com/BioInfoTools/BBMap<br>https://github.com/s-andrews/FastQC<br>https://github.com/MultiQC/MultiQC<br>https://github.com/andersen-lab/ivar<br>https://github.com/OpenGene/fastp<br>https://github.com/uwb-linux/prinseq<br>https://github.com/FelixKrueger/TrimGalore<br>https://github.com/usadellab/Trimmomatic<br>https://github.com/marcelm/cutadapt<br>https://github.com/swiftbiosciences/primerclip<br>https://github.com/kbaseapps/BBTools<br>https://github.com/tommyau/bamclipper |
| **Read mapping** | BWA-MEM<br>BWA-MEM2<br>Minimap2<br>Bowtie 2 | https://github.com/lh3/bwa<br>https://github.com/bwa-mem2/bwa-mem2<br>https://github.com/lh3/minimap2<br>https://github.com/BenLangmead/bowtie2 |
| **SNV calling** | iVAR<br>LoFreq<br>FreeBayes<br>BCFTools<br>V-PIPE<br>Nanopolish<br>VarScan<br>GATK<br>SAMtools<br>Medaka | https://github.com/andersen-lab/ivar<br>https://github.com/CSB5/lofreq<br>https://github.com/freebayes/freebayes<br>https://github.com/samtools/bcftools<br>https://github.com/cbg-ethz/V-pipe<br>https://github.com/jts/nanopolish<br>https://github.com/dkoboldt/varscan<br>https://github.com/broadinstitute/gatk<br>https://github.com/samtools/<br>https://github.com/nanoporetech/medaka |
| **Relative abundance estimation (LRA), allele frequency estimation (AF), lineage detection (LD)** | Freyja (LRA)<br>QualD (LD)<br>SAM Refiner (AF)<br>Kallisto-VLQ (LRA)<br>iVAR (AF) | https://github.com/andersen-lab/Freyja<br>https://gitlab.com/treangenlab/quaid<br>https://github.com/degregory/SAM_Refiner<br>https://github.com/baymlab/wastewater_analysis<br>https://github.com/andersen-lab/ivar |
| | VaQuERo (LRA)<br>Cojac (LD)<br>Lineagespot (AF)<br>LCS (LRA) | https://github.com/fabou-uobaf/VaQuERo<br>https://github.com/cbg-ethz/cojac<br>https://github.com/BiodataAnalysisGroup/lineagespot<br>https://github.com/rvalieris/LCS |
| ***De novo* assembly** | SOAPdenovo<br>Velvet<br>ABySS<br>IBDA<br>SPAdes<br>Edena<br>Vicuna<br>Haploflow<br>Flye | https://github.com/aquaskyline/SOAPdenovo2<br>https://github.com/dzerbino/velvet<br>https://github.com/bcgsc/abyss<br>https://github.com/loneknightpy/idba<br>https://github.com/ablab/spades<br>https://github.com/ddhz/edena<br>https://www.broadinstitute.org/viral-genomics/vicuna<br>https://github.com/hzi-bifo/Haploflow<br>https://github.com/fenderglass/Flye |
| **Lineage and clade assignment, Phylogeny** | IQtree2<br>Pangolin<br>NextClade<br>UShER | https://github.com/Cibiv/IQ-TREE<br>https://github.com/stevenlovegrove/Pangolin<br>https://github.com/nextstrain/nextclade<br>https://github.com/yatisht/usher |
| **Pipelines** | COVID-19 VIral Epidemiology Workflow (C-VIEW)<br>CFSAN Wastewater Analysis Pipeline (C-WAP)<br>PiGx SARS-CoV-2 Wastewater Sequencing Pipeline<br>VLQ-nf | https://github.com/ucsd-ccbb/C-VIEW<br>https://github.com/CFSAN-Biostatistics/C-WAP<br>https://github.com/BIMSBbioinfo/pigx_sars-cov-2<br>https://github.com/rki-mf1/vlq-nf |